\documentclass[a4paper,11pt]{article}    
\pdfoutput=1 
\usepackage{jheppub}
\usepackage{epsfig} 
\usepackage[T1]{fontenc} 
\def\arXiv#1{\href{http://arxiv.org/abs/#1}{arXiv:#1}}
\def\arXiv#1#2{\href{http://arxiv.org/abs/#1}{arXiv:#1}}

\def\doi#1#2{\href{http://doi.org/#1}{#2}}

\def\be{\begin{eqnarray}}
\def\ee{\end{eqnarray}}
\def\bea{\begin{eqnarray}}
\def\eea{\end{eqnarray}}
\newcommand{\nn}{\nonumber}
\newcommand\para{\paragraph{}}

\def\Dslash{\,\,{\raise.15ex\hbox{/}\mkern-12mu D}}
\def\Dbarslash{\,\,{\raise.15ex\hbox{/}\mkern-12mu {\bar D}}}
\def\delslash{\,\,{\raise.15ex\hbox{/}\mkern-9mu \partial}}
\def\delbarslash{\,\,{\raise.15ex\hbox{/}\mkern-9mu {\bar\partial}}}
\def\pslash{\,\,{\raise.15ex\hbox{/}\mkern-9mu p}}
\def\calDslash{\,\,{\raise.15ex\hbox{/}\mkern-12mu {\cal D}}}

\def\lae{\mathrel{\mathop{\smash{\lower .5 ex \hbox{$\stackrel<\sim$}}}}}
\def\lae{\mathrel{\mathop{\smash{\lower .5 ex \hbox{$\stackrel>\sim$}}}}}

\title{\boldmath Spatially modulated instabilities of holographic gauge-gravitational anomaly}

\author{Yan Liu$^{\textcolor{red}{a,\,b}}$ and Francisco Pena-Benitez$^{\textcolor{red}{c}}$} 

\affiliation{ $^{\textcolor{red}{a}}$Department of Space Science, and International Research Institute 
of Multidisciplinary Science, Beihang University,  Beijing 100191, China}
\affiliation{ $^{\textcolor{red}{b}}$Instituto de Fisica Teorica UAM/CSIC, 
Universidad Autonoma de Madrid,  Cantoblanco, 28049 Madrid, Spain}
\affiliation{ $^{\textcolor{red}{c}}$Dipartimento di Fisica, Universit\`a di Perugia, 
I.N.F.N. Sezione di Perugia, Via A. Pascoli, I-06123 Perugia, Italy}

\emailAdd{yanliu@buaa.edu.cn, benitez@pg.infn.it}

\abstract{We performed a study of the perturbative instabilities in Einstein-Maxwell-Chern-Simons theory with a gravitational Chern-Simons term, which is dual to a strongly coupled field theory with both chiral and mixed gauge-gravitational anomaly.  With an analysis of the fluctuations in the near horizon regime at zero temperature, we found that there might be two possible sources of instabilities. The first one corresponds to a real mass-squared which is below the BF bound of AdS$_2$, and it leads to the bell-curve phase diagram at finite temperature. The effect of mixed gauge-gravitational anomaly is emphasised. Another source of instability is independent of gauge Chern-Simons coupling and exists for any finite gravitational Chern-Simons coupling. There is a singular momentum close to which unstable mode appears. The possible implications of this singular momentum are discussed. 
Our analysis suggests that the theory with a gravitational Chern-Simons term around Reissner-Nordstr\"om black hole is unreliable 
unless the gravitational Chern-Simons coupling is treated as a small perturbative parameter. 
}

\begin{document}
\maketitle
\flushbottom
\pagestyle{plain} \setcounter{page}{1}
\newcounter{bean}
\baselineskip16pt



\section{Introduction}


\para It is well known that there is a novel spatially modulated instability for AdS Reissner-Nordstr\"om (RN) black hole in five dimensional Einstein-Maxwell theory with a Chern-Simons term \cite{Nakamura:2009tf, {Ooguri:2010kt}, {Donos:2012wi}}. More precisely, for a large enough Chern-Simons coupling constant, 
below some critical temperature the translational invariance in one of the spatial directions of RN black hole is spontaneously  broken and a spatially modulated black hole with helical current is the preferred state. From the perspective of gauge/gravity duality \cite{Zaanen:2015oix}, one concludes that the dual field theory, namely a four dimensional strongly coupled chiral anomalous system, suffers from spatially modulated instability. Thus this holographic model may provide interesting dual descriptions of systems in nature, and clarify several aspects of their phase structures, including quark-gluon plasma and Weyl/Dirac (semi-)metals.
\para On the other hand, in a four dimensional chiral field theory system, besides pure gauge anomalies, 
there is another type of anomaly, the mixed gauge-gravitational anomaly, which is the gravitational contribution to the chiral anomaly (see e.g. \cite{AlvarezGaume:1983ig}). These two types of anomaly are distinguished depending on the three insertions in the triangle diagram being of spin one, or an insertion of spin one and the other two being of spin two. For an anomalous system at finite temperature and chemical potential, the dynamics of anomaly enters into the energy-momentum conservation and current  conservation equations and will contribute to anomaly-induced transports \cite{Son:2009tf,{Jensen:2012kj}}.  Remarkably, the mixed gauge-gravitational anomaly leads to important observable effects in many body physics at finite temperature including e.g. chiral vortical effect \cite{Landsteiner:2011cp}, odd (Hall) viscosity \cite{Landsteiner:2016stv} and negative thermal magnetoresistivity \cite{Lucas:2016omy}. 
\para However, so far a study on possible effects of spatially modulated instability due to the mixed gauge-gravitational anomaly is missing. In the holographic context the mixed gauge-gravitational anomaly is encoded in gravity via a gravitational Chern-Simons term \cite{Landsteiner:2011iq}. Unlike the pure gauge Chern-Simons term, the gravitational one is higher order in derivatives endowing the theory with the subtleties of higher derivative gravity, we will come back to this point in section \ref{sec2}. Our aim is to explore the possible 
instability due to the mixed gauge-gravitational anomaly of an  anomalous system. Our strategy to study the instability is by analysing the fluctuation around Reissner-Nordstr\"om black hole solution to examine the possible unstable modes. 
\para At zero temperature we studied the fluctuations around the near horizon geometry AdS$_2$ $\times$ R$^3$ and analyse the instabilities. From the AdS$_2$ point of view, we found two sources of instabilities. One is the Breitenlohner-Freedman (BF) bound violation which is essentially the same as the case without mixed gauge-gravitational anomaly. Another one is related to a degenerate point in the equations, which is characterized by a momentum that we will call singular momentum. At finite temperature the first source of instability plays as a sufficient condition for the bell-curve phase diagram. The second source of instability exists at any finite gravitational Chern-Simons coupling. 
There always exists a mode with BF bound violation close to the singular momentum, and we could expect the existence of a spontaneous symmetry breaking solution. We will discuss the implications of this singular momentum for finite temperature in section \ref{sec:singular}.
\para The rest of the paper is organised as follows. In section \ref{sec2} we outline the setup of gravitational theory and fix the conventions of the paper. In section \ref{sec3} we study the perturbative instabilities for the RN black hole on the gravity side to construct the phase diagram as well as discuss possible physical implications of a singular momentum. We conclude in section \ref{sec5} with a summary of results and a list of open questions.

\section{Setup}
\label{sec2}
\para Let us first briefly review the basic setup for the holographic systems encoding both the gauge and mixed gauge-gravitational anomaly in the dual field theory \cite{Landsteiner:2011iq}.  The minimal setup of five dimensional gravitational action we consider is 
\bea\label{eq:action}
\mathcal{S}&=&\frac{1}{2\kappa^2}\int d^5x\sqrt{-g}\bigg[R+\frac{12}{L^2}-\frac{L^2}{4}F^2+\epsilon^{\mu\nu\rho\sigma\tau}A_\mu\Big(\frac{\alpha L^3}{3} F_{\nu\rho}F_{\sigma\tau}+\lambda L^3 R^\beta_{~\delta\nu\rho}R^\delta_{~\beta\sigma\tau}\Big)\bigg]~~~~~~
\eea
where $2\kappa^2$ is five-dimentional gravitational coupling constant,  $L$ is the AdS radius and $F_{\mu\nu}=\partial_\mu A_\nu-\partial_\nu A_\mu$ is the gauge field strength.  In the following we will set $2\kappa^2=L=1$. Note that $\alpha$ and $\lambda$ are dimensionless quantity. 
\para The equations of motion for this system are\footnote{We set the Levi-Civita tensor $\epsilon_{\mu\nu\rho\nu\beta}=\sqrt{-g}\varepsilon_{\mu\nu\rho\nu\beta}$ with $\varepsilon_{0123r}=1$.} 
\bea\label{eq:eom1}
R_{\mu\nu}-\frac{1}{2}g_{\mu\nu}\Big(R+12 -\frac{1}{4}F^2\Big)
-\frac{1}{2} F_{\mu\rho}F_{\nu}^{~\rho}
-2\lambda \epsilon_{\alpha\beta\rho\tau(\mu|} \nabla_\delta (F^{\beta\alpha} R^{\delta~~\rho\tau}_{~|\nu)})&=&0\,,
\\
\label{eq:eom2}
\nabla_\nu F^{\nu\mu}+\epsilon^{\mu\tau\beta\rho\sigma} \Big[\alpha F_{\tau\beta}F_{\rho\sigma}
+\lambda R^{\delta}_{~\xi\tau\beta}R^\xi_{~\delta\rho\sigma}\Big]&=&0\,,
\eea 
where $A_{(\mu}B_{\nu)}=\frac{1}{2}(A_\mu B_\nu+A_\nu B_\mu).$
\para The Reissner-Nordstr\"om black brane is a solution of the system
\be\label{eq-bg-rn}
ds^2=-U dt^2+\frac{dr^2}{U}+r^2(dx^2+dy^2+dz^2)\,,~~A= \phi dt
\ee
where 
\be U= r^2-\frac{r_+^4}{r^2}+\frac{\mu^2}{3}\bigg(\frac{r_+^4}{r^4}-\frac{r_+^2}{r^2}\bigg)\,, ~~~\phi=\mu\Big(1-\frac{r_+^2}{r^2}\Big)\,. \ee
From the AdS/CFT correspondence \cite{Zaanen:2015oix}, the isotropic dual field theory lives at the conformal boundary $r\to \infty$ with the temperature 
\be T=\frac{U'(r_+)}{4\pi}=\frac{r_+}{\pi}-\frac{\mu^2}{6\pi r_+}\ee 
and a finite (axial) chemical potential $\mu$. The Chern-Simons terms do not contribute to the thermodynamical quantities for Reissner-Nordstr\"om black brane. The pure gauge anomaly and mixed gauge-gravitational anomaly of this system are fixed by $\alpha$ and $\lambda$ respectively \cite{Landsteiner:2011iq}.  
More precisely, the dual field theory describes a chiral fluid with an anomalous axial U(1) symmetry whose conservation equations in terms of covariant currents are 
\be
\nabla_a T^{ab}&=& F^{ba}J_a-
2\lambda \nabla_b(\varepsilon^{cefl}F_{ce}R^{ba}_{~~fl})
\,,\\
\nabla_a J^{a}&=& -\big(\alpha\varepsilon_{abcd}F^{ab}F^{cd}+\lambda\varepsilon^{abcd}R^e_{~fab}R^f_{~ecd}\big)\,.
\ee
\para 
In the context of the low energy effective field theory, other higher derivative terms should, in principle, be included in theory (\ref{eq:action}) and the four-derivatives terms should be treated as perturbative corrections. However, 
the particular theory (\ref{eq:action}), which is considered from a bottom-up holographic point of view\footnote{Our perspective is different from the low energy effective field theory context, in which all the possible terms with the same order in derivatives should be included.} has the advantage of being the minimal model from which the dual (non)-conservation equations precisely match the ones of chiral fluid. Therefore, to  capture the physical effects due to the anomalies, including anomaly-induced transport and phase transitions, (\ref{eq:action}) is sufficient for simplicity. 

\para There are reasons to treat the four-derivatives term non-perturbatively.\footnote{The use of this ``non-perturbative''  approach has been investigated in the holographic context for various other higher derivative theories, e.g. \cite{Brigante:2008gz, {Myers:2010pk}}.} In the pure AdS$_5$ case, the mixed anomaly term does not contribute to the linear fluctuations of the theory (\ref{eq:action}), and no ghost will appear, this means that one could treat $\lambda$ as an arbitrary constant. In the Schwartzschild black hole case, as we shown in the appendix \ref{app:a} from the perspective of quasi-normal modes, no unstable modes seems to appear when treating $\lambda$ as an order one number. Moreover, for a single Weyl fermion, we have $\alpha\sim\lambda\sim \mathcal{O}(1)$, i.e. the mixed gravitational anomaly has the same importance of the chiral anomaly, and we remind the reader our motivation is purely phenomenological. From these perspectives, it is tempting to treat $\lambda$ as a non-perturbative parameter  although in general the higher derivative terms might have potential causality issues \cite{Camanho:2014apa}\footnote{So far, this theory has been extensively explored in the description of several effects related with (mixed) anomalies \cite{Landsteiner:2016stv, Landsteiner:2011iq,Bhattacharyya:2016knk, Chapman:2012my}.}, and suffer of Ostrogradsky's instabilities.\footnote{We will not consider these aspects here, given the linearised equations of motion are second order.}

\para In the following, we shall study the effects due to ``non-perturbative'' gravitational Chern-Simons coupling on the stability of Reissner-Nordstr\"om black hole.

\section{Perturbative instability at zero and finite temperature}
\label{sec3}
\para In this section, we will study the perturbative instability of the gravitational system at zero and finite temperature respectively. The effect of the mixed gauge-gravitational anomaly on the instability will be explored.

\subsection{Instability at zero temperature}
\para In this subsection we will study the stability of the extremal RN black hole solution in the gravitational theory (\ref{eq:action}). 
\para At zero temperature, i.e. extremal RN ($r_+=\mu/\sqrt{6}$), the near horizon geometry of RN black hole\footnote{Notice that $\rho=\frac{1}{12}(r-r_+)^{-1}$. We also set $r_+=1$, i.e. we are in unit of $\mu=\sqrt{6}$.
} is  
AdS$_2$ $\times$ R$^3$ 
\bea\label{eq:ads2}
ds^2=\frac{1}{12 \rho^2}(-dt^2+d\rho^2)+(dx^2+dy^2+dz^2)\,,~~~A_t=\frac{1}{\sqrt{6}\rho}\,.
\eea
and the BF bound for the scalar mass squared in AdS$_2$ is $-3.$
\para Following \cite{Nakamura:2009tf}, we consider fluctuations around the near horizon geometry (\ref{eq:ads2})
\be\label{zeroTfluc}
\delta g_{ai}=h_{ai}(t, \rho) e^{i kx}\,,~~~\delta A_i=a_i (t, \rho) e^{ik x}
\ee
with $i\in\{y, z\}$ and $a\in\{t, x\}$ around the near horizon geometry (\ref{eq:ads2}).
By plugging these fluctuations into the equations of motion (\ref{eq:eom1}, \ref{eq:eom2}), we find that the equations from  gravity sector are
\bea
-6\rho^2 h_{ti}''-12 \rho h_{ti}'+\frac{k^2}{2}h_{ti}+\frac{ik}{2}\partial_t h_{xi}+\sqrt{6} a_i'~~~~~~~&&\nn\\
\label{eq:g1}+i4\sqrt{6}k\lambda \epsilon^{ij}\big(12\rho^2 h_{tj}''
+24 \rho h_{tj}'-k^2 h_{tj}-ik\partial_t h_{xj}+2\sqrt{6} a_j'\big)&=&0 \,,\\
\label{eq:g2}
h_{xi}''-\partial_t^2 h_{xi}+ik\partial_t h_{ti}-i 8\sqrt{6}k\lambda \epsilon^{ij}\big(h_{xj}''-\partial_t^2 h_{xj}+ik\partial_t h_{tj}\big)&=&0\,, \\
\label{eq:g3}
\sqrt{6}i\partial_t a_i-6i \rho^2\partial_t h_{ti}'-\frac{k}{2} h_{xi}'+i4\sqrt{6}k\lambda \epsilon^{ij} \big(k h_{xj}'+12i\rho^2\partial_t h_{tj}+2\sqrt{6}i\partial_t a_j\big)&=&0 
\eea
where the prime $'$ is the derivative with respect to $\rho$ and $\epsilon^{yz}=-\epsilon^{zy}=1$. Note that only two of them are independant since $-i\partial_t (\ref{eq:g1})+\frac{k}{2} (\ref{eq:g2})+\partial_\rho(\ref{eq:g3})=0$. Thus we shall focus on (\ref{eq:g1}) and (\ref{eq:g3}).
The equation from gauge sector is 
\be\label{eq:ga1}
12\rho^2(a_i''-\partial_t^2 a_i)-k^2 a_i-24\sqrt{6}\rho^2\big(h_{ti}' +8\sqrt{6} i k\lambda \epsilon^{ij}h_{tj}'\big)-16\sqrt{6}ik\alpha\epsilon^{ij} a_j=0\,.
\ee
\para Via the sequence of field redefinitions
\bea
\varphi_i &=& a_i + i8\sqrt{6} k\lambda\epsilon^{ij} a_j - \sqrt{6} \rho^2\big (h_{ti}' - i8\sqrt {6} k\lambda\epsilon^{ij} h_ {tj}'\big)\,,\\
\label{eq:red}
  \Phi_a &=& (a_y + i a_z, a_y - ia_z , \varphi_y + i\varphi_z , \varphi_y - i\varphi_z)\,,
   \eea 
(\ref{eq:g3}) can be rewritten as two  first order PDEs
\be\label{eq:newg3}
\partial_t\Phi_{a} + 4ik\left((-1)^ak\lambda + \frac{1}{8\sqrt{6}}  \right)\left(h'_{xy} -i(-1)^a  h'_{xz}\right)=0\,, \quad a=3,4
\ee
from which we know that in general the dynamics of $h_{xi}$ is totally determined by $\Phi_{3,4}.$ 
\para The redefinition (\ref{eq:red}) has a special point $k=\pm k_s$ with
\be\label{eq_ks} 
k_s=\frac{1}{8\sqrt{6}\lambda}\,,
\ee
 such that the four new redefined fields are not independent. At this specific value redefinition (\ref{eq:red}) reduces to
\bea\label{eq:degen}
\Phi_{3}=2\Phi_{1} \quad \text{for}\quad k=k_s\,, \text{~~~(or~~} \Phi_{4}=2\Phi_{2} \quad\text{for} \quad k=-k_s)\,.
\eea
We will refer to it as the singular momentum, for reasons that will be clarified below. 
\para In order to study the possibility of violating the BF bound, we will analyze first the system at the singular momentum, and then at arbitrary values of $k$.  From (\ref{eq:newg3}) we have $\partial_t\Phi_{3}=0$ or $\partial_t\Phi_{4}=0$ respectively. 
 The equations for the independent dynamical fields are\footnote{In the following equation until (\ref{eq:ins1}) the first lower index corresponds to the case with $k=k_s$ and the second lower index is for $k=-k_s$. Either the first index or the second index is chosen in these equations. }  
\bea\label{eq:demass1}
\Box_{\text{AdS}_2}\Phi_{4,3} - \frac{1}{384 \lambda ^2}\Phi_{4,3} &=&0 \,,\\
\label{eq:demass2}
\Box_{\text{AdS}_2}\Phi_{2,1}  - \left(\frac{1}{384 \lambda ^2}-\frac{2 \alpha }{\lambda }\right) \Phi_{2,1} &=&0\,,
\eea
where $\Box_{\text{AdS}_2}=12 \rho^2 (-\partial_t^2+\partial_\rho^2)$. The other two dependent (nondynamical) degrees of freedom are determined via the relations
\bea \label{eq:phi12}
\Phi_{1,2}'= \partial_t \Phi_{1,2}&=&0\,,\\ 
\label{eq:phi12-2}
 48\sqrt{ 6} \rho^2 h_{t\pm}'+ \left(\frac{2 \alpha }{\lambda }+\frac{1}{384 \lambda ^2}\right) \Phi_{1,2} &=&0\,,
\eea
with $h_{t\pm}=h_{ty}\pm i h_{tz}$.\footnote{ Eq. (\ref{eq:phi12-2}) should be understood  taking $h_{t+}$ (or $h_{t-}$) with $\Phi_1$ (or $\Phi_2$).} 
%
\para Thus we only have two dynamical field $\Phi_{4,3}$ and $\Phi_{2,1}$ at this singular momentum. Furthermore, we observe that the system is automatically diagonal in these dynamical variables, and more important is the fact the $\Phi_{4,3}$ has a positive mass, therefore always above the BF bound, however the field $\Phi_{2,1}$ has a mass  that can be negative depending on the values of $(\alpha,\lambda)$. The region of instability is given by the inequality
\be\label{eq:ins1}
|\alpha| > \Big{|}\frac{3\lambda}{2}+\frac{1}{768 \lambda}\Big{|} \quad  \textrm{and} \quad  \alpha\lambda >0\,.
\ee
Thus exactly at the singular momentum there is no instability if $|\alpha|<\frac{1}{8\sqrt{2}}.$
\para Now we will discus the case $k\neq \pm k_s$, in this case we recover the original four degrees of freedom and the equations they satisfy read
\be\label{eq:AdS2}
(\Box_{\text{AdS}_2}-\mathbb{M}_\pm^2){\bf \Phi}_\pm=0\,,
\ee
with ${\bf \Phi}_+=\left(
\begin{array}{c}
 \Phi_{1}\\
 \Phi_{3} \\
\end{array}
\right) $, ${\bf \Phi}_-=\left(
\begin{array}{c}
 \Phi_{2}\\
 \Phi_{4} \\
\end{array}
\right) $ and 
\be
\label{eq_Mpm}
 \mathbb{M}_\pm^2=\left(
\begin{array}{cc}
 k^2-72  \pm16 \sqrt{6} k (\alpha - 12 \lambda ) + \frac{96 }{1 \mp 8\sqrt{6} k \lambda }
 \quad& 24 \left(1 - \frac{2}{1 \mp 8 \sqrt{6} k \lambda }\right) \\
 -k^2 \left(1 \pm 8 \sqrt{6} k \lambda \right) & k^2 \\
\end{array}
\right) \,.\ee
\para For generic $\alpha$ and $\lambda$, the eigenvalues\footnote{We emphasise that the eigenvalues do not depend on the field redefinition of $\Phi$, e.g. instead of (\ref{eq:red}) one would redefine $\Psi_a=(a_y + i a_z, a_y - ia_z , r^2(h_{ty} + ih_{tz}) , r^2(h_{ty} - ih_{tz}))$, however, as long as $k\neq \pm k_s$ we will get the same mass eigenvalues.} are
\bea\label{eq:masssquare}
m_{j\pm}^2&=&k^2-36\pm 8\sqrt{6}k(\alpha-12\lambda)+\frac{48k_s }{k_s\mp k}+ \\
&& (-1)^j\sqrt{\bigg(-36\pm8\sqrt{6}k(\alpha-12\lambda)+\frac{48k_s }{k_s\mp k}\bigg)^2
+24 \frac{k^2}{k_s}\frac{(k_s\pm k)^2}{k_s \mp k}} \nn
\eea
with $j=1,2.$ 
\para There is a lot of information in the above formulae:
\begin{itemize}
\item In absence of the anomaly, i.e. $\alpha=\lambda=0$, we have ${m_j^2}_+={m_j^2}_-$ ($j=1,2$) and $\text{min} (m_1^2, m_2^2)=24-2\sqrt{6}$ which is above the BF bound of (\ref{eq:ads2}). This is the well-known fact that there is no spatially modulated  instability for RN black hole in Einstein-Maxwell theory. 

\item For $\lambda =0$, we only have gauge anomaly and we have 
\bea
m_{j\pm}^2 =k^2+ 12  \pm 8\sqrt{6}k\alpha+(-1)^j 2 \sqrt{6} \sqrt{k^2+(\sqrt{6}\pm 4\alpha k)^2}\,, 
\eea
which violate the BF bound when $|\alpha|>\alpha_c\approx 0.1448$ as studied for the first time in \cite{Nakamura:2009tf}. For $\alpha=\alpha_c$, the minimal value of mass square is at $k_v/\mu\approx 1.5349$.  
Thus it indicates a spatially modulated phase transition. 

\item When $k\to k_s$ (or $k\to -k_s$), either $m_{1+}^2$ or $m_{2+}^2$ (or $m_{1-}^2$ or $m_{2-}^2$ ) become negative/positive infinity\footnote{This limit depends on from which side of $\pm k_s$ it is approached. The limit has a discontinuity at $k=\pm k_s$, i.e. from one side it is $\pm\infty$ while from the other side is zero. } while the other mass squares coincide with (\ref{eq:demass1}, \ref{eq:demass2}). 
This is due to the degenerate definition of fields at $k=\pm k_s$ in (\ref{eq:red}) which leads to (\ref{eq:degen}). 
Because of this artificial infinity in (\ref{eq:masssquare}), we called $k_s$ singular momentum. 
Nevertheless, when $k\neq \pm k_s$, (\ref{eq:masssquare}) applies. 
Thus we conclude that close to $\pm k_s$ BF bound is always violated and it is a source of instability in the system.

\item  $m_j^2$ is invariant under $(k, \alpha, \lambda)\to -(k,\alpha, \lambda)$.   This means that in order to study the unstable modes in momentum space, we can simply focus on the case with one of the $k, \alpha, \lambda$ parameter space to be positive.

\item The square root  in (\ref{eq:masssquare}) can become complex and this is totally different from the case without the mixed gauge-gravitational anomaly \cite{Nakamura:2009tf}, if that happens the system will be unstable because the unitarity condition $m_j^2>m_\text{BF}^2$ will not be satisfied. We label the critical momentum beyond which the square root become complex as $k_*$. Note that $k_*$ is a function of $\alpha$ and $\lambda$ and it can be obtained by solving the equation
\be\label{eq_kstar}
\bigg(-36 + 8\sqrt{6}k_*(\alpha-12\lambda)+\frac{48k_s }{k_s - k_*}\bigg)^2
+24 \frac{k_*^2}{k_s}\frac{(k_s + k_*)^2}{k_s  - k_*}=0\,.
\ee
If $k_*(\alpha,\lambda)$ is the solution for $m_{j+}^2$, $k_*(-\alpha,-\lambda)$ will be for $m_{j-}^2$. 
The number of solutions\footnote{In this item, we focus on the positive momentum case.} for $k_*$ of (\ref{eq_kstar}) depend on $\alpha$ and $\lambda$. There are two situations. One could have a unique solution $k_*$ and for $k_*<k$ the mass square become complex. Another case is that one may have three solutions $k_{*1}<k_{*2}<k_{*3}$. For $k_{*1}<k<k_{*2}$ or $k_{*3}<k$, the mass square is complex while for $k_{*2}<k<k_{*3}$ the mass square is real. We label the smallest $k_{*1}$ as $k_*$.  It can be proved that
\be\label{eq:relksks}
 k_s<|k_*| 
\ee  
independent of the value of $\alpha$.  
  
It is more apparent to see the above statements in the limit $k\to\infty$, we have  
\be\label{eq:impres1}
m_{j\pm}^2=k^2+(-1)^j 8(54)^{1/4}\sqrt{\mp\lambda}k^{3/2}+\dots,~~~(j=1,2)\,,
\ee
thus one of the mass square becomes complex when $\lambda\neq 0$. 
This indicates that for a non-zero value of the mixed anomaly coupling $\lambda$ and arbitrary value of Chern-Simons coupling $\alpha$, the theory should be {\em unstable} around RN black hole.\footnote{This reminds us the recent study  \cite{Bhattacharyya:2016knk,{Camanho:2014apa}} based on positivity of relative entropy in pure AdS background for the same system without gauge anomaly. However, their conclusion on causality violation at arbitrary finite value of gravitational Chern-Simons theory 
is for flat Minkowski spacetime. It is necessary to examine the issue of causality in AdS case. Note that for zero density system we do not find any perturbative instability modes similar to (\ref{eq:impres1}).}

\item For $\alpha=0$, the complex mass momentum $k_*$ can be solved to be
\bea\label{eqn:forkstar}
\nonumber \pm 3k_sk_* &=& 6 + k_s^2 + \frac{\left(k_s^4 + 48 k_s^2 + 36 \right)}{\left( \left(k_s^4 + 153 k_s^2 + 432 + 9 k_s \sqrt{2 k_s^4 + 213 k_s^2 + 144} \right) k_s^2 + 216 \right)^{1/3} } +   \\
&&\left(\left(k_s^4+153 k_s^2+432+9 k_s \sqrt{2 k_s^4+213 k_s^2+144}\right) k_s^2+216\right)^{1/3}\,.
\eea
This complicated function has a simple behaviour in the two following limits
 \bea
 k_* &=& \pm 6k_s^{-1} \,, \quad k_s\to 0  ~~~~(\text{or~} \lambda\to \infty) \\
 k_* &=& \pm k_s\,, ~~~\quad k_s\to \infty  ~ ~~(\text{or~} \lambda\to 0)\,.
 \eea
For generic $\alpha$, we show a numerical loglogplot in Fig. \ref{fig_kstar} and one observe that when $k_s\to 0$ or $k_s\to\infty$ we have the similar behaviour as the case $\alpha=0.$

\begin{figure}[h!]
\begin{center}
\includegraphics[width=0.6\textwidth]{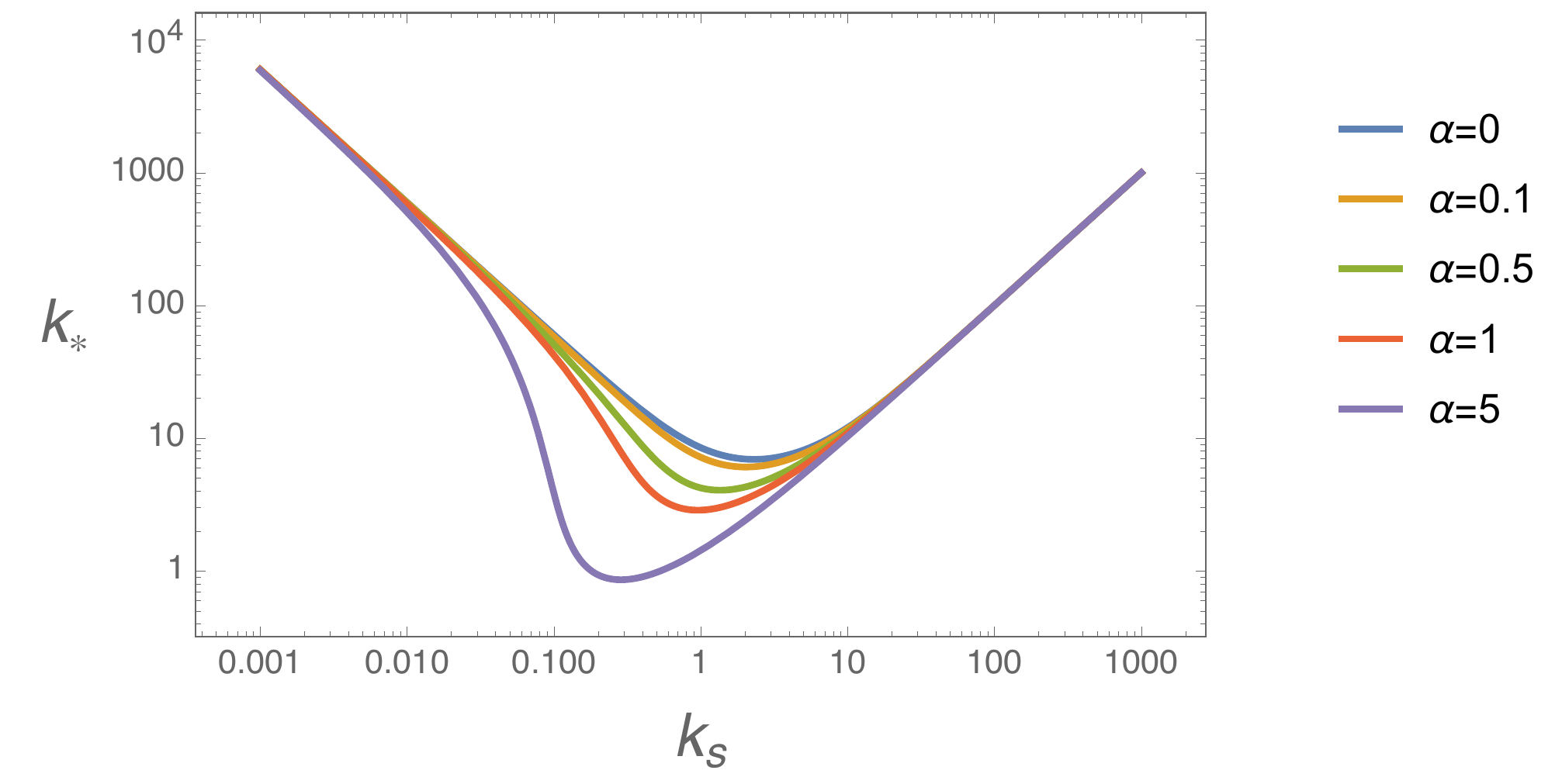}
\caption{\small The loglogplot of $k_*$ as a function of $k_s$ at different values of $\alpha$. The behaviour of $k_*$ at small $k_s$ or large $k_s$ is independent of $\alpha$. Note that one can always find a solution of $k_*$ at arbitrary $\alpha,\lambda$ and this means that the instability always exists whenever $\lambda\neq 0.$}
\label{fig_kstar}
\end{center}
\end{figure}

\item When $|k|< |k_*|$, the mass squares in (\ref{eq:masssquare}) are real. We have $m_1^2< m_2^2$ and we can focus on the $m_1^2$ sector to explore the instability. 
For fixed $\alpha$ and $\lambda$, violating the BF bound (i.e. $\text{min} \{m_{j\pm}^2\}<-3$) will give a finite regime of $k$ in which there is an instability which is a generalisation of \cite{Nakamura:2009tf}. After minimising (\ref{eq:masssquare}) with respect to momentum, we obtain the location of the minima $k_{\textrm{min}}(\alpha,\lambda)$ as a function of the Chern-Simons couplings, plugin back this function into the mass function (\ref{eq:masssquare}) and equating to the BF bound we obtain one of the Chern-Simons coupling in terms of the other, this function determines the critical values of the Chern-Simons couplings. In Fig. \ref{fig_valcrit} we show the dependance of the critical Chern-Simons coupling $\alpha_c$ as a function of $\lambda$. For fixing $\lambda$, when $\alpha>\alpha_c$ the BF bound is violated. Interestingly, we found with finite $\lambda$, the critical $\alpha$ could be smaller than the gauged SUGRA bound in which $\alpha=1/(4\sqrt{3})\simeq 0.1443...>{\alpha_c}_{\text{min}}$. However, the complete action for SUGRA up to this order involves other higher derivative terms \cite{Cremonini:2008tw}, it would be interesting to study the spatially modulated instability in the SUGRA.\footnote{See \cite{Takeuchi:2011uk} for an attempt in this direction. 
}

\begin{figure}[h!]
\begin{center}
\includegraphics[width=0.56\textwidth]{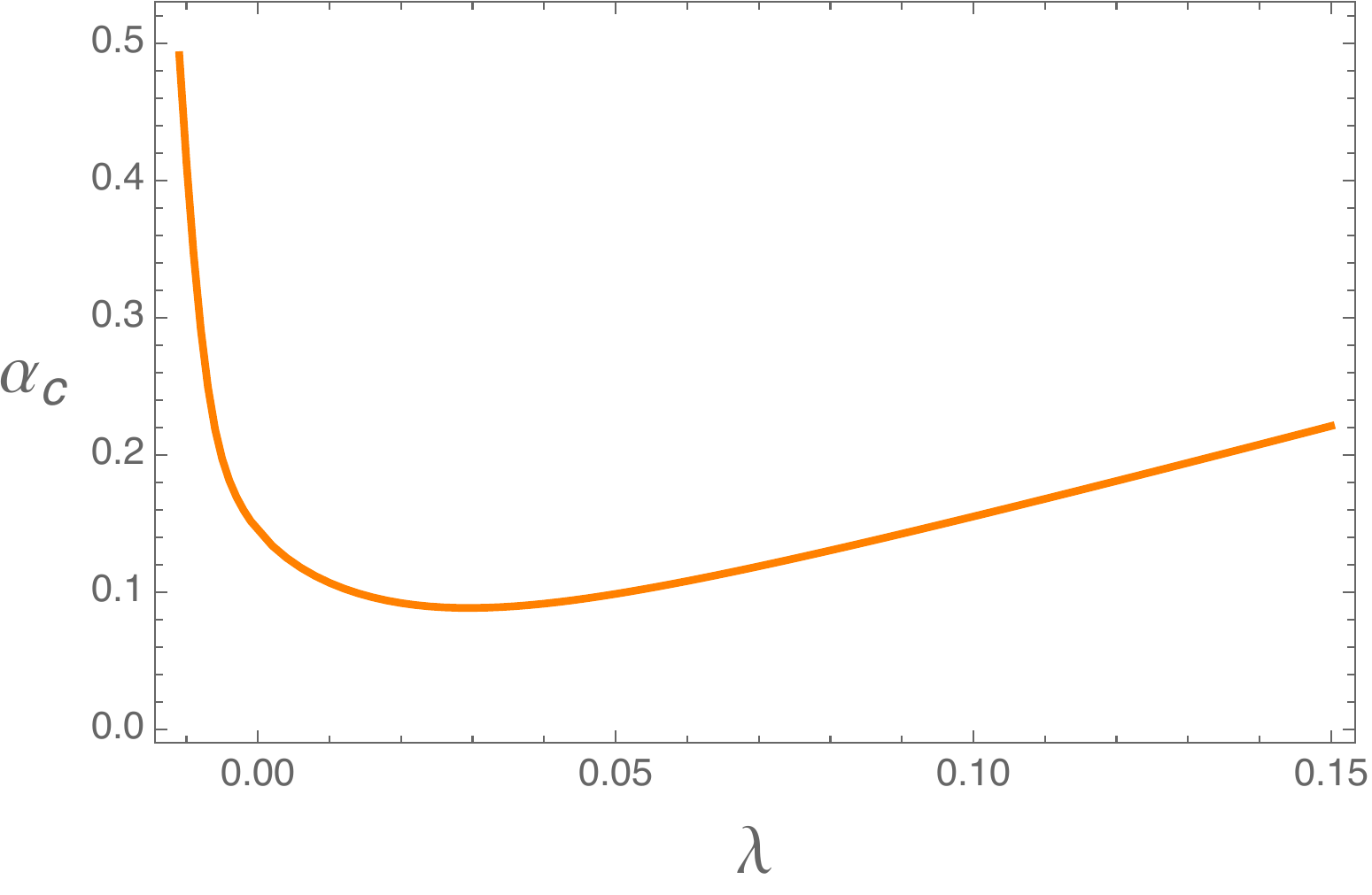}
\caption{\small  The critical Chern-Simons coupling $\alpha_c$ as a function of $\lambda$ for the local minimum in the mass square being equal to the BF bound. Note that ${\alpha_c}_{\text{min}}=0.088$ at $\lambda=0.029.$ For fixed $\lambda$, when $\alpha_c<\alpha$, the minimal mass square is below the BF bound at value $k<k_s.$
}
\label{fig_valcrit}
\end{center}
\end{figure}

\item Now there might be three different momentum scales, $k_s$ defined in (\ref{eq_ks}), $k_*$ in (\ref{eq_kstar}) and $k_v$ at which the BF bound is violated. These three scales may lead to an instability of the system for fixed $\alpha$ and $\lambda$. Since the reation (\ref{eq:relksks}) is always satisfied and the BF bound is always violated when $k>k_s$, we will not consider $k_*$ from now on. 

We can summarise as follows. The gravitational system around RN AdS black hole should have an instability as long as $\lambda\neq 0$. When $\alpha<\alpha_c$, the sources of instabilities are from $k_s$. When  $\alpha_c<\alpha$,  the sources of instabilities are from $k_s$ and also $k_v$.  

Let us illustrate the instabilities with the behaviour of $m_{1-}^2$ which is shown in Fig. \ref{fig_new}.  In both plots  regions green and black refer to violation of BF bound, in the green region $m_{1-}^2<-3$, whereas in the black region $m_{1-}^2\in \mathcal C$ . When $\alpha$ takes values in the interval $0\leq \alpha<{\alpha_c}_{\text{min}}$ the instability is tuned by $k_s$, and for $\alpha> {\alpha_c}_{\text{min}}$ a new instability island (red region) appears in the middle of the stable (white) region.   When $\alpha=0.21$,  its behaviour is shown in the right plot. As similar to the the case $\alpha=0$ we always have instabilities from $k_s$ and $k_*$, while we also have a new instability island (red region) in which the mass square is below the BF bound. If we further increase $\alpha$, the unstable island become even bigger and will cross the $k=-k_s$ curve.

\begin{figure}[h!]
\begin{center}
\includegraphics[width=.45\textwidth]{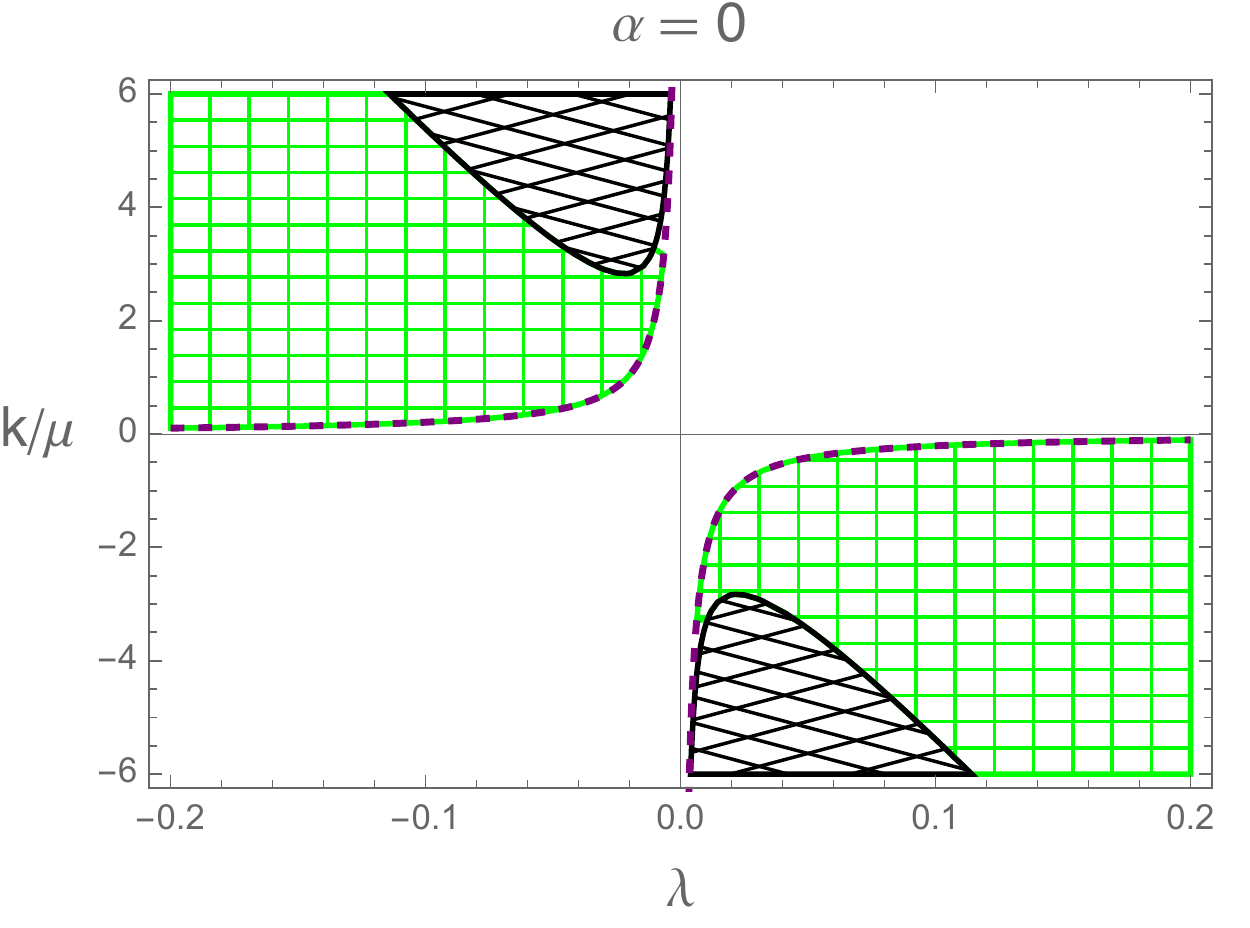}
\includegraphics[width=.45\textwidth]{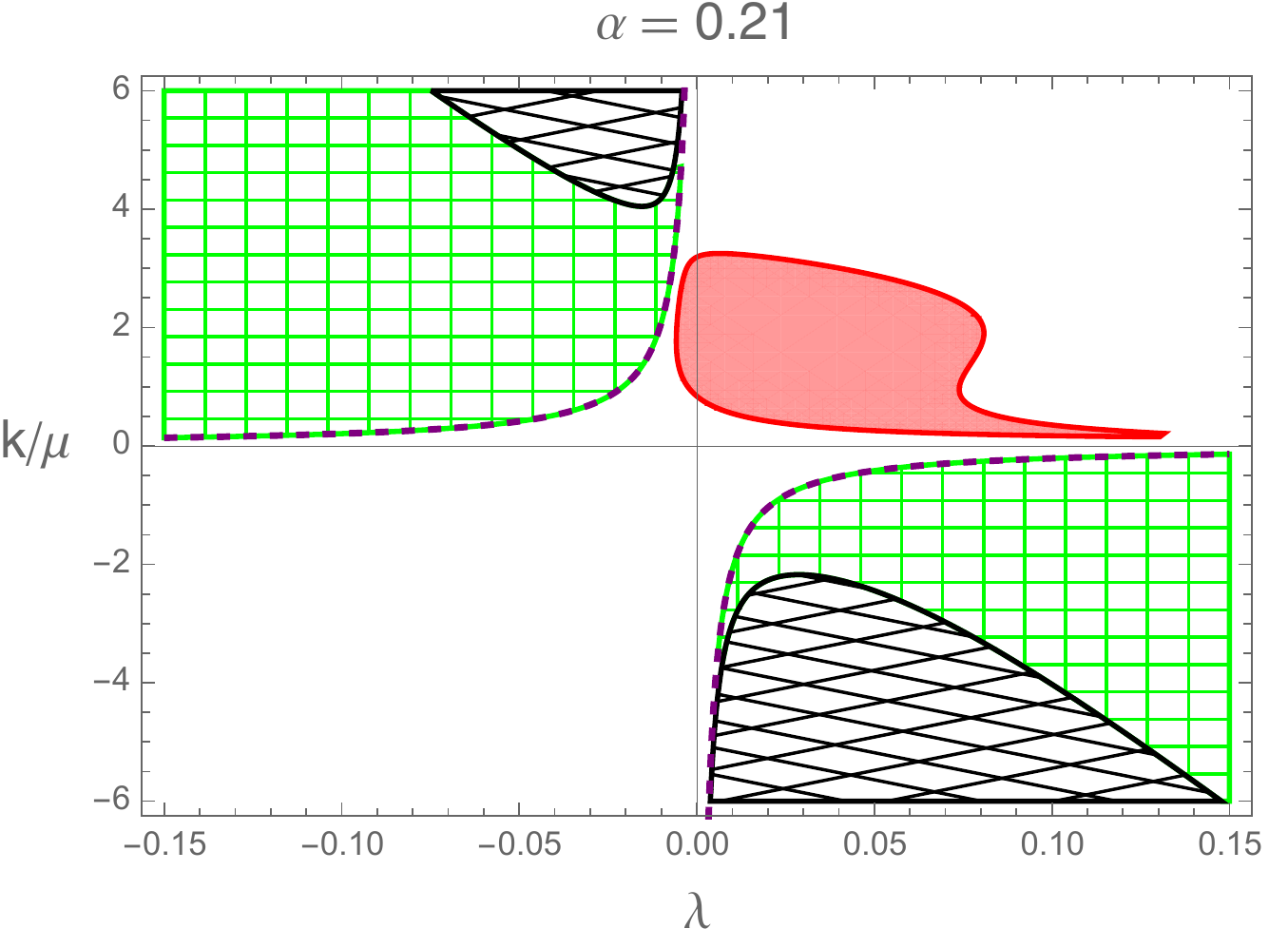}
\caption{\small  The behaviour of $m_{1-}^2$ as a function of $k/\mu,\lambda$, 
at $\alpha=0$ (left) and $\alpha=0.21$ (right). In both plots, $m_{1-}^2$ is complex inside the black meshed region, below the BF bound in the green meshed region and above the BF bound in the rest of the white space. 
The dashed purple line corresponds with $k=-k_s$ in (\ref{eq_ks}) and the green region are the instabilities inherited from the scale $k_s$. In the right plot 
$m_{1-}^2$ is also below the BF bound inside the red region (related to scale $k_v$) and this is related to gauge Chern-Simons term $\alpha$. The sector of $m_{1+}^2$ can be obtained by the transformation $k\to-k$. Since for real mass squares we have $m_{1\pm}^2\leq m_{2\pm}^2$, it is sufficient to study $m_1^2$ for the instability issue. 
}
\label{fig_new}
\end{center}
\end{figure}

\end{itemize}

\subsection{Instability and phase diagram}
\begin{figure}[h!]
\begin{center}
\includegraphics[width=0.49\textwidth]{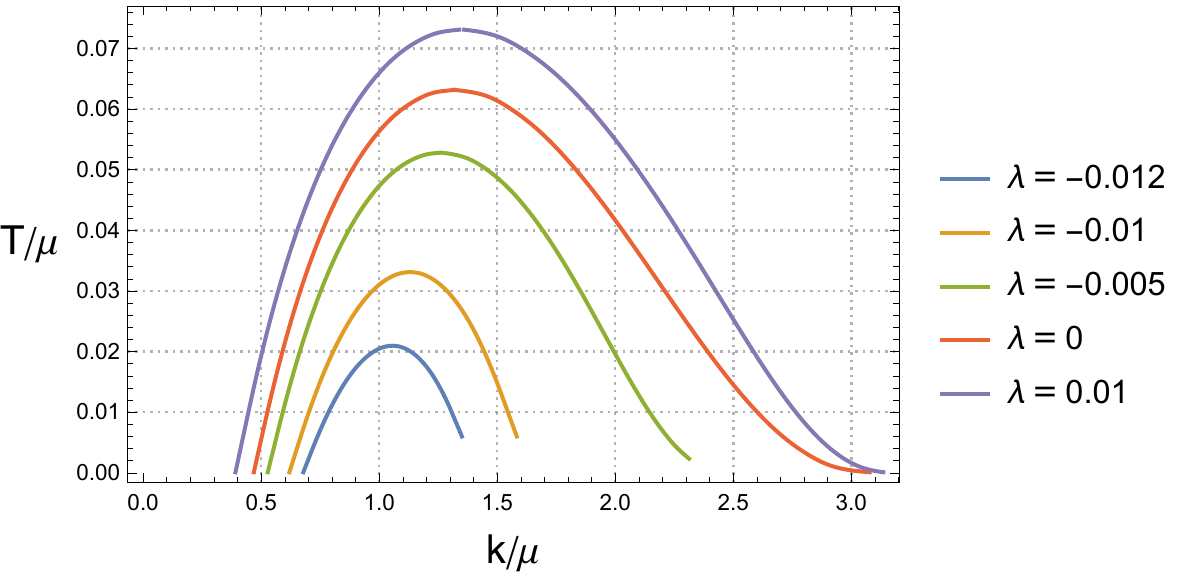}
\includegraphics[width=0.49\textwidth]{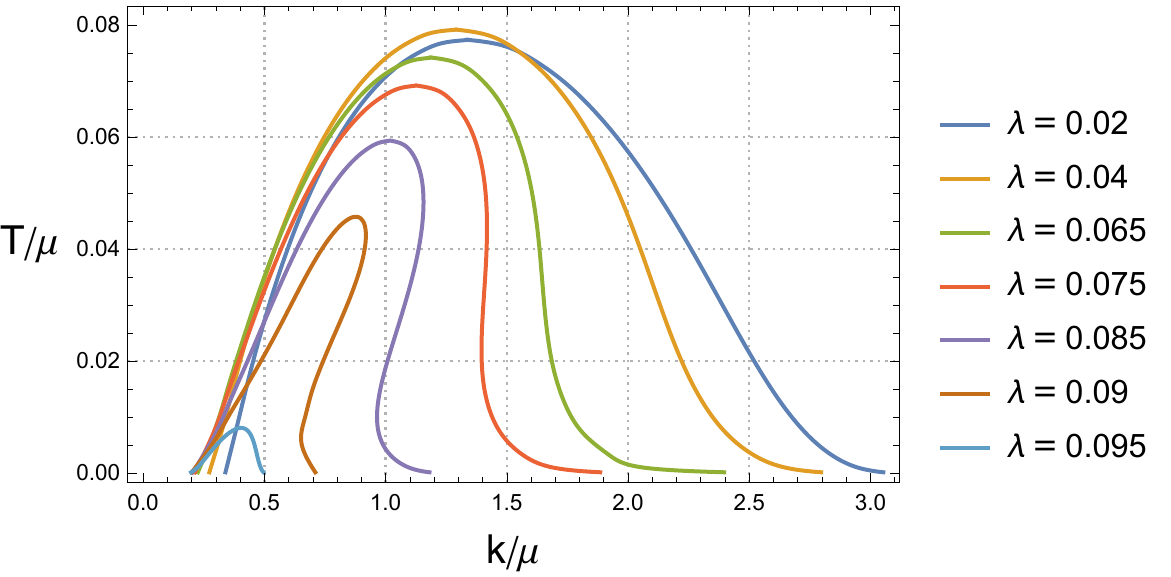}
\caption{\small The ``bell curve'' phase diagram at $\alpha\simeq 0.213$ (i.e. $\alpha/\alpha_c(\lambda=0)=1.47$) for different values of  $\lambda$. Note that when $\lambda= -0.01, -0.012$, the minimal temperature on the right side of bell-curve can not approach zero. }
\label{fig:phase1}
\end{center}
\end{figure}

\para In the previous subsection, we studied the instability of the near horizon region of the extremal RN solution and we found the system has two different sources of instability. The first is intimate ligated with having $\alpha >\alpha_c$ and it will be the source of the typical bell curves that were found in the literature \cite{Nakamura:2009tf}, the second exists if $\lambda \neq 0$ and essentially it is the BF bound violation when $k\to \pm k_s$.  In this subsection, we will study the linear perturbation around the full Reissner-Nordstr\"om black hole in AdS$_5$ at both finite and zero temperature to study the effect of mixed gauge-gravitational anomaly on the stability of RN black hole. We will also comment on the nature of $k_s$. 
\para Similar to the zero temperature near horizon analysis, we turn on the fluctuations (\ref{zeroTfluc}) to study their equations in the full spacetime. We focus on the zero frequency limit to look for static solutions. It turns out we have $(a_y+ia_z, h_{ty}+ih_{tz})$ sector decoupling from $(a_y-ia_z, h_{ty}-ih_{tz})$ sector. Thus we consider the following static fluctuations around RN black hole geometry  background (\ref{eq-bg-rn}) \cite{Donos:2012wi}\footnote{These fluctuations correspond to the sector $(\Phi_2, \Phi_4)$ in (\ref{eq:AdS2}) and the final phase diagram is expected to be consistent with Fig. \ref{fig_new}.}
\be\label{eq:flufiniteT}
\delta (ds^2)=2 Q dt \omega_2\,,~~~\delta A= a \omega_2
\ee
with the helical 1-forms of Bianchi VII$_0$ with pitch $k$
\be\label{eq:basis}
\omega_1=dx\,,~~~\omega_2=\cos(kx) dy+\sin(kx) dz\,,~~~\omega_3=\sin(kx) dy-\cos(kx) dz\,.
\ee
From the ansatz above,  it is clear that the translational symmetry is preserved along $y$- and $z$- directions while broken along the $x$-direction. The residual symmetry is so-called  helical symmetry, i.e. a one-parameter family with  a translation in $x$- direction combined with rotation in ($y,z$) plane,   
$x\to x-\epsilon$, $(y,z)\to (y,z)+\epsilon k (z, -y)$. 
\para After substituting (\ref{eq:flufiniteT}) into the equations of motion (\ref{eq:eom1}, \ref{eq:eom2}) we obtain\footnote{We will resctric our analysis to $k>0$ taking advantage of the invariance of the system under $(k,\alpha,\lambda)\to-(k,\alpha,\lambda)$.}
\bea\label{eq:finiteT}
\Big[(r+4k\lambda\phi')Q'\Big]'+\Big(-2r(k^2+4 U)-8k\lambda (k^2-4U)\phi'\Big)\frac{Q}{2 r^2 U}~~~~~&&\\+\big(r\phi'-\lambda(48k-\frac{24kU'}{r}+2k\phi'^2)\big)a'-4k\lambda U''' a&=&0\,,\nn\\
a''+\Big(\frac{U'}{U}+\frac{1}{r}\Big)a'-\big(k-8\alpha r \phi'\big)\frac{k}{r^2 U}a+\Big(r^3\phi'-4k\lambda(r^2U''-2r U'+2U)\Big)\frac{rQ'-2Q}{r^4 U}&=&0\,.\nn
\eea
First of all we observe that the system has the usual singularities at the boundary, horizon and a possible new singularity when $r_s+4k\lambda\phi'=0$ has a real solution.\footnote{It indicates the presence of a characteristic hypersurface in the system of ODE's \cite{Reall:2014pwa}.} In the case of having a real solution if $r_s$ is in between the horizon and the bulk,  
there will be subtleties to solve the system. Thus in the search of static normalisable solutions we will restrict the analysis to the cases when new singularity is behind the horizon $r_s<r_+$ or it does not exist. Later we will discuss the physical origin of this singularity and its possible implications in the full phase diagram of the theory.
\para Now let us analyse the near horizon and near boundary behaviours of the fields in (\ref{eq:finiteT}). At finite temperature, near horizon $r\to r_+$, we have 
\be\label{eq:nhfiniteT}
Q=c_1 (r-r_+)+c_2 (r-r_+)^2+\dots\,,~~~~
a=d_0+d_1 (r-r_+)\dots.
\ee
with 
\be
d_1= \frac{1}{r_+^2T}\left[\left( k^2 - 16 \alpha  k \mu  \right)\frac{d_0}{4\pi} + \left(48 k \lambda   - \mu  \right)\frac{c_1 r_+}{2\pi} -36T k \lambda  c_1\right]
\ee
and 
\bea
\nn c_2 &=& \frac{1}{4\pi Tr_+^2(r_+^2+8 k \lambda  \mu )}\left[\bigg( 4 \lambda  k^3 \mu +k^2 \Big(\frac{r_+^2}{2}+1152 \lambda ^2 (2 r_+-3 \pi  T)^2\Big)+ \right.\\
\nn && \left.~~~~~~~~ 48 \lambda  k \mu  r_+ (7 \pi  T-4 r_+)+2 r_+^2 \big(\mu ^2-\pi  r_+ T\big) \bigg) c_1 \right. + \\
\nn && \left.~~\Big(24 \lambda  k^3 (2 r_+-3 \pi  T) - k^2 (\mu  r_+  + 384 \alpha  \lambda  \mu  (2 r_+-3 \pi  T))+ \right.\\
\label{eq:c2}&& \left.~~~~~~~~k \big(16 \alpha  \mu ^2 r_++192 \pi  \lambda  r_+ T (8 \pi  T-7 r_+)\big)\Big)d_0\Big]\right. \,.
\eea
Notice that from Eq. (\ref{eq:c2}) it is necessary to avoid the point $ r_+^2+8 k \lambda  \mu=0$, which corresponds precisely with the finite temperature redefinition of the singular momentum $k=-k_s$, where
\be 
\label{eq:ksfinT}
k_s(\lambda, T)=\frac{1}{288\mu\lambda} \left(3 \pi  T+\sqrt{6\mu^2+9 \pi ^2 T^2 }\right)^2.
\ee 
 This singular point appears as a consequence of the degeneration of the system at the characteristic surface $r_s+4k\lambda\phi'=0$ in Eq. (\ref{eq:finiteT}). 
 In order to solve the system  at this point, it would be necessary to find a new near horizon expansion. We leave the analysis of this case to the next subsection. 
\para Near the boundary $r\to \infty$, we have
\bea\label{eq:nbnd}
\nn Q&=&Q_0 r^2-\frac{k^2}{4}Q_0+\frac{k^4 Q_0}{16}\frac{\ln r}{r^2}+\frac{Q_2}{r^2}+\ldots\nn\,,\\
a&=&a_0-\frac{k^2 a_0}{2}\frac{\ln r}{r^2}+\frac{a_2}{r^2}+\ldots\,.
\eea
We are interested in the normalisible solutions which signal the onset of instability, i.e. we look for solution with $a_0=Q_0=0.$
Thanks to the scaling properties of the background, we can always set $\mu=1$, the parameters $T$ and $k$ at which the static solution exists will be totally determined by the condition $a_0=Q_0=0.$  
A convenient numerical method for seeking this static solution is the double shooting method. For details on the method, see e.g. \cite{{Krikun:2013iha},{Kiritsis:2015hoa}, {Andrade:2015iyf}}. The idea is as follows, one construct  two independent solutions from the horizon to some matching point $r_m$ using the horizon data ($c_1,d_0$). Then we shoot from the boundary to $r_m$ another pair of independent solutions using the boundary parameters $(a_2, Q_2)$ with $a_0=Q_0=0$. If the static solution exists, there will be a smooth connection between solutions at the matching point $r_m$, which is equivalent to the condition of vanishing the Wronskian  at this point. 
\para Starting from the normal state, i.e. RN black hole solution, we study the onset of $T/\mu$ at which static solution exists as a function of $k/\mu$ for different $(\alpha, \lambda)$. As shown in Fig. \ref{fig:phase1}, we have the ``bell-curve'' phase diagrams of $T/\mu$ as a function of $k/\mu$  for finite $\alpha\simeq 0.213$  and different values of $\lambda$.  The ``bell-curve'' behavior in the literature \cite{Nakamura:2009tf} is not qualitatively modified by the mixed gauge-gravitational anomaly. Actually we find the bell-curves are consistent with the BF bound violation region at zero temperature, as shown in the red region of the right plot of Fig. \ref{fig_new}. When we  increase $\lambda$ from negative values the red island (see Fig. \ref{fig_new}) becomes wider very fast, similar as we observe in left plot of Fig. \ref{fig:phase1}. On the other hand, the upper momentum shows a turning point in Fig. \ref{fig_new} at some positive $\lambda$, such that the island becomes narrower. We observe a similar behavior in the right plot of  Fig. \ref{fig:phase1}.
We also point out that the bell curves at $\lambda= -0.01, -0.012$ do not end at $T=0$, due to the fact that unstable region intersects $-k_s$ (Eq. \ref{eq:ksfinT}), avoiding convergence of the numerics. 

\begin{figure}[h!]
\begin{center}
\includegraphics[width=0.45\textwidth]{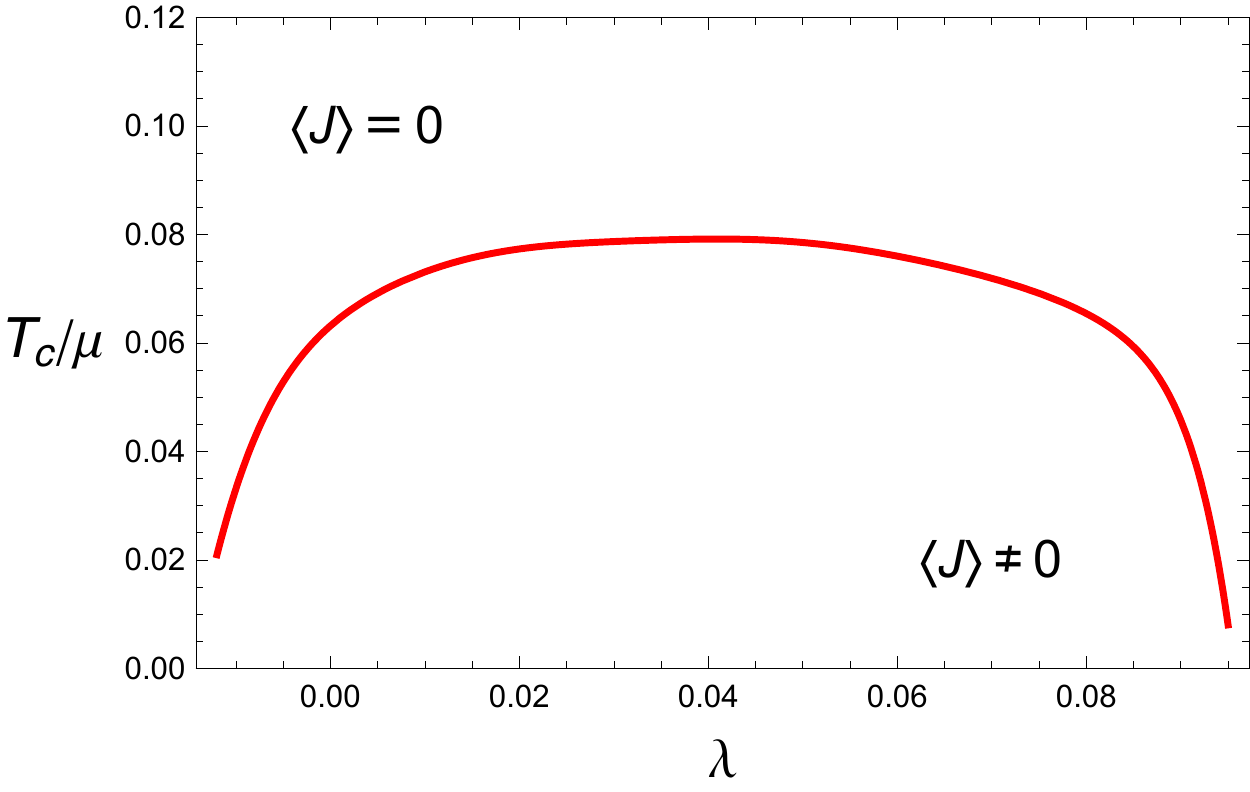}
\includegraphics[width=0.45\textwidth]{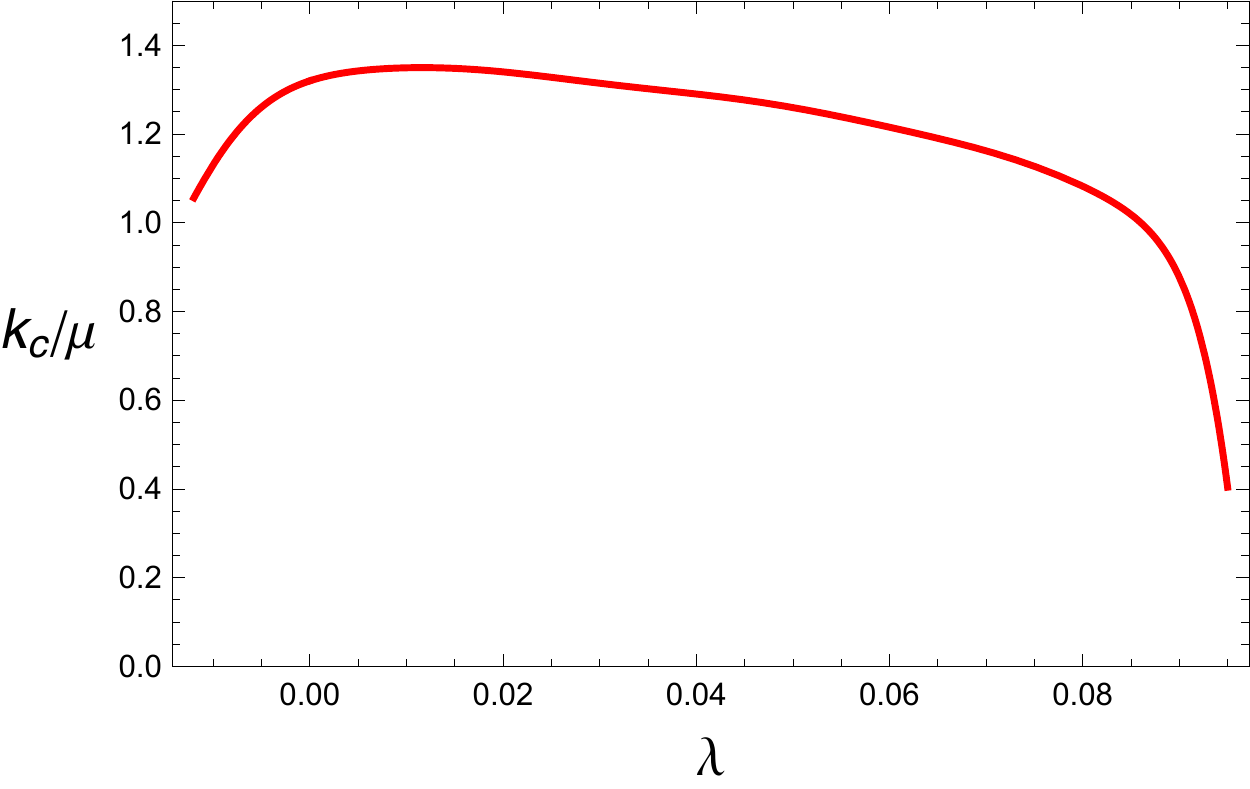}
\caption{\small The critical temperature $T_c/\mu$ ({\em left}) which is the onset of the spatially modulated phase transition and the corresponding critical momentum $k_c/\mu$ ({\em right}) for different $\lambda$ when $\alpha\simeq 0.213$.}
\label{fig:Tc_lambda}
\end{center}
\end{figure}

%
\para The static solution with the highest critical temperature $T_c$ which happens at $k_c$, corresponds to the onset of the spatially modulated phase.  The behaviour of $T_c$ and $k_c$ depending on $\lambda$ can be found in Fig. \ref{fig:Tc_lambda}. When $\lambda<0$, $T_c$ decreases, however the behaviour is not monotonic when $\lambda>0$. The corresponding $k_c/\mu$ behaves similarly as shown in the right plot.  Hence, in the parameter space without possible singularity 
one can increase $\lambda$ or decrease $\lambda$ to tune $T_c$ to be zero. At the onset of this (possible) quantum phase transition, the pitch $k_c$ is nonzero.  

\begin{figure}[h!]
\begin{center}
\includegraphics[width=0.56\textwidth]{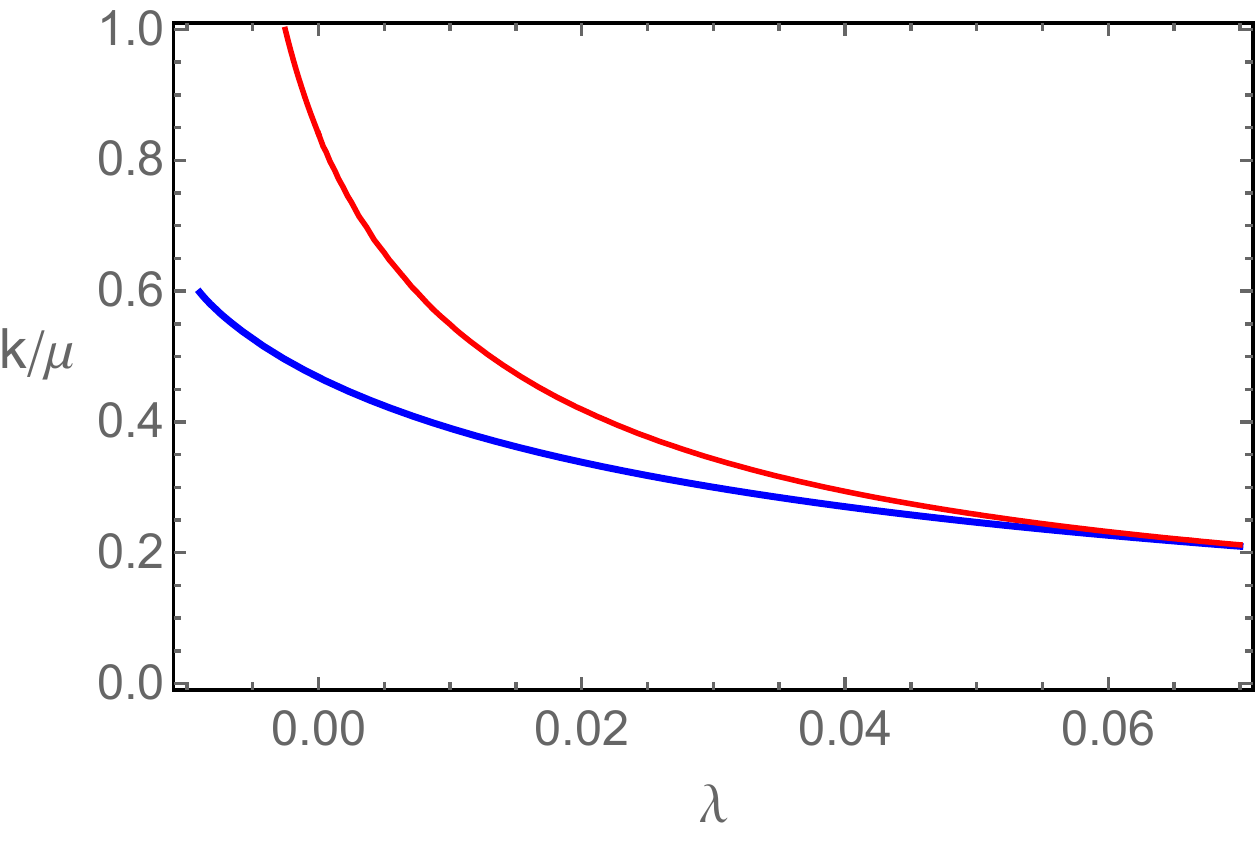}
\caption{\small The blue line is the lowest critical value of momentum as a function of $\lambda$ (when $\alpha\simeq 0.213$) at which the static solution appears. It corresponds to the lower boundary of the bell curve in Fig. \ref{fig:phase1} at zero temperature. The red line is the boundary of $m^2_{1-}$ at which the BF bound is saturated.}
\label{fig:zeroTss}
\end{center}
\end{figure}

%
\para For the case without mixed gauge-gravitational anomaly it was proven in \cite{Nakamura:2009tf} that the BF bound violation at zero temperature is a sufficient while not necessary condition for RN black hole to be unstable. To check the effect of the mixed anomaly we will construct static normalisable solutions at zero temperature. To do so, we build a near horizon expansion at zero temperature to be used in the numerical integration towards the boundary. 
The expansion reads  
\be\label{eq:nh0T}
Q=q_1 (r-r_+)^{\beta_1}+q_2 (r-r_+)^{\beta_2}\,,~~~~~a=b_1 (r-r_+)^{\beta_1-1}+b_2 (r-r_+)^{\beta_2-1}
\ee
with 
\bea
b_j &=&\frac{2\beta_j r_+ (48k\lambda+\sqrt{6} r_+)}{-12 r_+^2 \beta_j(\beta_j-1)+k^2+16\sqrt{6}kr_+\alpha} q_j \,,\\
\beta_j &=&\frac{1}{2}+ \frac{\sqrt{3}}{6}\bigg(15+\frac{k^2}{r_+^2}+\frac{8\sqrt{6}k}{r_+}(\alpha-12\lambda)+
\frac{384\sqrt{6}k\lambda}{r_+-8\sqrt{6}k\lambda}+\nn\\
&&\frac{(-1)^j}{r_+^2}\sqrt{\bigg(-36r_+^2+8\sqrt{6}kr_+(\alpha-12\lambda)+\frac{48 r_+^3 }{r_+-8\sqrt{6} k\lambda}\bigg)^2
+24 k^2 r_+\frac{(r_++8\sqrt{6}k\lambda)^2}{r_+-8\sqrt{6}k\lambda}} ~\Bigg)^{1/2}\,.\nn
\eea
In units of $r_+=1$, i.e. $\mu=\sqrt{6}$, we have $\beta_j=\frac{1}{2}+\frac{\sqrt{3}}{6}\sqrt{3+m^2_{j-}}$ with $m^2_{j-}$ defined in (\ref{eq:masssquare}). 
With this new near horizon condition and using the double shooting method, we find that static solutions with $k\leq k_v$ exist, which is shown in Fig.  \ref{fig:zeroTss}. In this plot we compare the momentum at which the BF bound is violated (red curve) with the lowest momentum found at which the static normalisable solution exists (blue curve). Note that when momentum takes value at the blue curve, it is above the BF bound (at least for $\lambda<0.06$). Thus we confirm for the system with the mixed gauge-gravitational anomaly, that BF violation at $T=0$ is a sufficient while not necessary condition for the instability.

\subsubsection{Comments on the singularity and instability}
\label{sec:singular}
%
\para Now let us comment on the other source of instability, $k_s$ at finite temperature. From (\ref{eq:finiteT}) it is clear  that $\lambda$ introduces a new ``singular'' point in the system at which the coefficient in front of $Q''$ vanishes.  Besides the standard poles at the horizon ($r=r_+$) and the boundary ($r=\infty$),  the new singular point appears as the solution of $r_s+4 k \lambda \phi '=0$, with the only real positive root being
\be\label{eq:rs}
r_s =\frac{1}{\sqrt{3}}(-2k \lambda\mu )^{1/4}\sqrt{3 \pi  T+ \sqrt{9 \pi ^2 T^2+6\mu^2}} \,,\quad \,k \lambda\mu<0\,. 
\ee
\para There are two possibilities to avoid having a singularity in the bulk, either $k \lambda\mu>0$ ($r_s$ not real) or hiding the singular point behind the horizon ($r_s<r_+$), which imply
\be\label{eq:contraintrs}
 -\frac{1}{288} \left(3 \pi  T+\sqrt{6\mu^2+9 \pi ^2 T^2 }\right)^2<k \lambda\mu \,,
\ee
which at zero temperature reduces to $ -\frac{1}{48}<\lambda k/\mu.$ Notice that the lower bound in (\ref{eq:contraintrs}) corresponds to $-k_s\lambda\mu$ with $k_s$ defined in Eq. (\ref{eq:ksfinT}). That gives the clear interpretation for $k_s$, as the momentum at which $r_s=r_+$.
\para Now let us make a generic discussion on the implication of the singularity. If we write an effective reduced action for fluctuations (\ref{eq:flufiniteT}), we have 
 \be \label{eq:efflang}
 \mathcal{L}_\text{eff}\supset (r+4k\lambda\phi') QQ''+\ldots.
 \ee 
When moving from the boundary to the horizon, Eq. (\ref{eq:efflang}) indicates that the kinetic term for $Q$ changes sign at $r=r_s$,\footnote{This $r_s$ is different from the ``accessible singularity'' found in \cite{Lippert:2014jma}. } suggesting therefore the existence of ghost-like modes.\footnote{We emphasize that at zero density the system is free of any type of instability (see appendix \ref{app:a}). Hence the calculations of transport coefficients such as chiral vortical conductivity \cite{Landsteiner:2011iq} and odd viscosity \cite{Landsteiner:2016stv} from this gravitational theory are reliable for this case.} 
\para Let us remind the reader that the full non-linear system is third order, that implies the holographic dictionary for the model should be properly studied and defined in detail. Also notice that the linearised problem around Reissner-Nordstr\"om is second order. If for example the charged black hole was a stationary axisymmetric-like, instead of a planar static one, the linear equations would be third order. On a background of this form, the singular point could disappear because the structure and existence of the characteristic surface is given by the coefficient of the highest derivative term in the equations. Of course, in this case a new problem emerge because the holographic dictionary needs to be understood.

\para From a top down point of view  $\lambda$ should be small and treated as a perturbative parameter, which would solve the problem.
For example, in $\mathcal{N}=4$ SYM, the chiral anomaly is of order $N_c^2$ while the mixed gauge-gravitational anomaly is zero, however, one might also attempt to add some flavors degrees of freedom to make the mixed gauge-gravitational anomaly $\lambda$ of order $N_c N_f$ \cite{Kimura:2011ef}. Then we would have $\lambda/N_c^2\sim N_f/N_c\ll 1$.  In the units $2\kappa^2=L=1$ that we use, this means the $\lambda$ term in (\ref{eq:efflang}) can be ignored. Nonetheless, our perspective is bottom up. Therefore, to better understand the physics of the singular surface we will study the case $r_s=r_+$ (i.e. $k=-k_s$), where no problematic sign in the kinetic term of the effective Lagrangian appears. The finite temperature near horizon expansion (\ref{eq:nhfiniteT}) has to be modified as follows
\bea\label{eq:nhspecial}
Q &=& (r-r_+)^{\gamma_1}(q_0+...)+c_0+c_1 (r-r_+)+\dots\,,\nn\\ 
a&=& (r-r_+)^{\gamma_1}(a_0+...)+d_0+d_1 (r-r_+)+\dots\,,
\eea
with 
\be
\gamma_1 = \frac{1}{2}\sqrt{\frac{3}{2}}\frac{\big{|}6r_+^2-5\mu^2\big{|}}{\sqrt{6\mu^2r_+^2-\mu^4}}\,, ~~~~
a_0 = \frac{\text{sign}(6r_+^2-5\mu^2)\sqrt{6}\,q_0}{\sqrt{6r_+^2-\mu^2}}\,,
\ee
where  
 $(c_i, d_i)$ ($i=1,2,...$)  are totally fixed by $(c_0, d_0)$. This expansion shows two problems, the first one is the presence of three integration constants ($q_0,c_0,d_0$), making the system undetermined. 
The second is even worst, each coefficients $(c_i, d_i)$ become singular when $T/\mu\simeq 0.05, 0.08, 0.14$, $0.36$, $0.53$, $0.72\ldots$ respectively.
\para In principle it is possible to understand our fluctuations around Reissner-Nordstr\"om as a special limit of the fully backreacted ansatz 
 \bea
 \label{eq:backreact}ds^2&=&-u f dt^2+\frac{dr^2}{u}+h \omega_1^2+r^2 e^{2\gamma} (\omega_2+Q dt)^2 +r^2 e^{-2\gamma} \omega_3^2\\
\label{eq:backreact2}  A&=& \phi dt+a \omega_2
\eea
with $\omega_i$ $(i=1,2,3)$ defined in (\ref{eq:basis}). At the black hole horizon $u(r_+)=Q(r_+)=0$ have to be satisfied, which suggest the condition $c_0=0$ in the near horizon expansion (\ref{eq:nhspecial}), 
however we still have the problem of the singularities in the coefficients $(c_i, d_i)$. Anyhow after trying to find static solutions avoiding the singularities, we realized that 
no normalisable solution can be constructed.

\begin{figure}[t!]
\begin{center}
\includegraphics[width=0.49\textwidth]{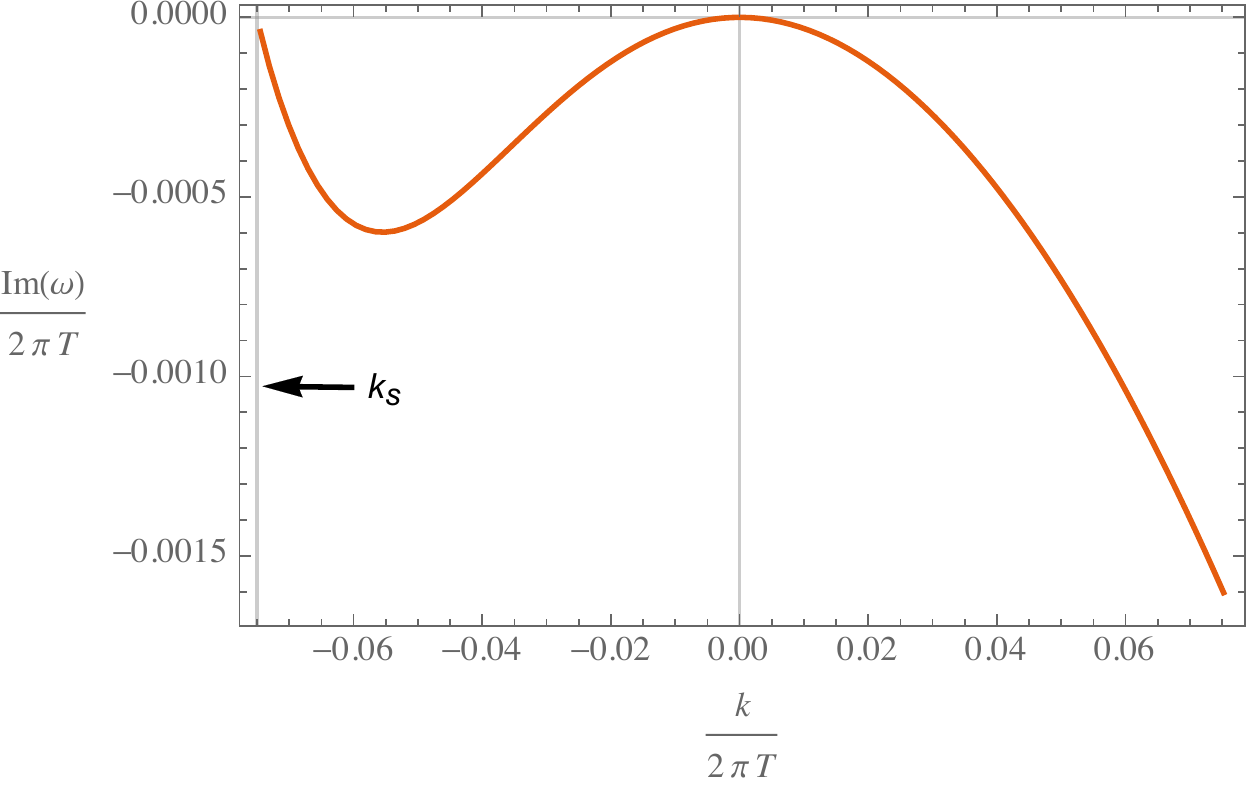}
\includegraphics[width=0.47\textwidth]{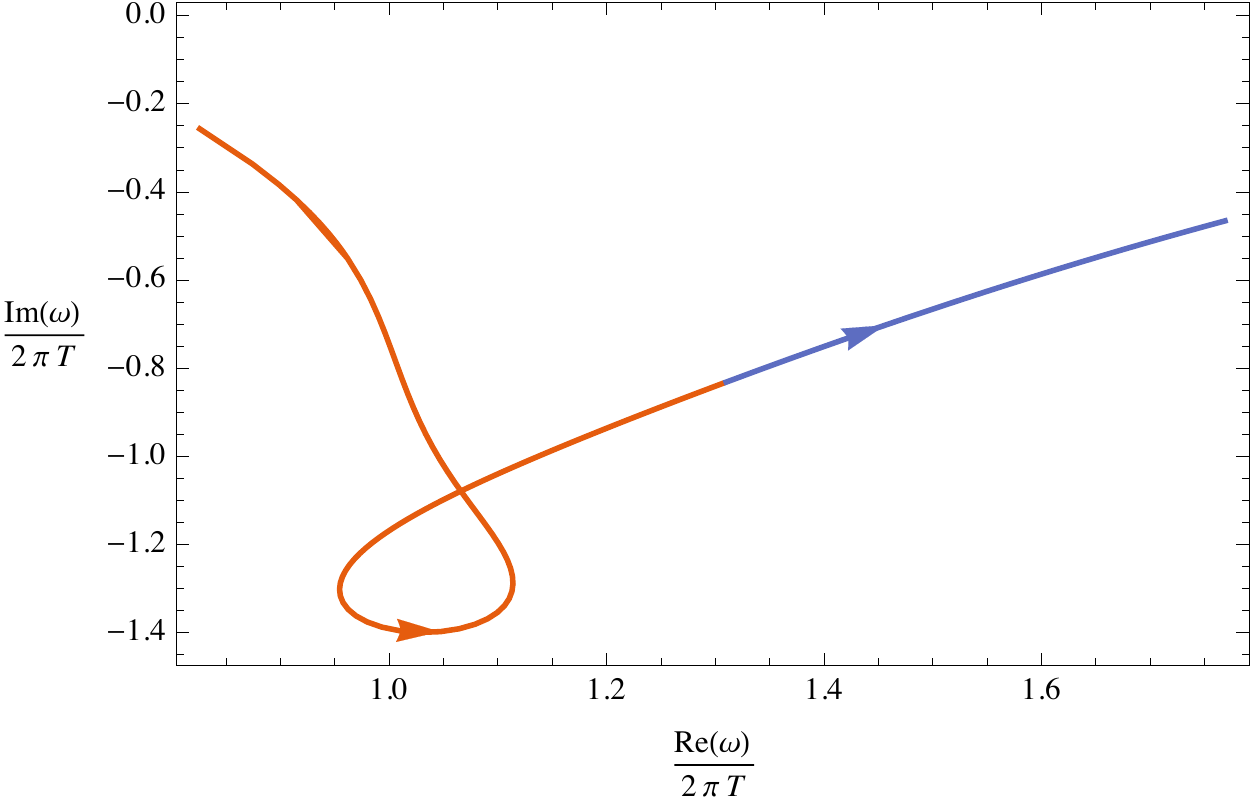}
\caption{\small Left: Lowest purely imaginary quasinormal mode as a function of momentum. Right: Complex frequency plane for the second quasinormal mode parametrized by the momentum, blue and red lines correspond to positive and negative momentum respectively. The arrows indicate the direction the momentum grows or decreases. The singular momentum is $k_s/(2\pi T)= -0.075$. The parameters are fixed as $\mu=1$, $\lambda=1$, $\alpha=0$ and $r_+=1$.}
\label{fig:qnmfinitemu}
\end{center}
\end{figure}

\para Given the unreliability of the linear problem when $r_s=r_+$, we will proceed to compute the quasinormal modes for $r_s\lesssim r_+$. Considering the presence of the characteristic surface is insensitive to the values of $\alpha$, as we observe in Fig. \ref{fig_new}, we will set it to zero. In Fig. \ref{fig:qnmfinitemu} we show the lowest quasinormal modes  after fixing the parameters as follow, $\mu=1$, $\lambda=1$ and $r_+=1$. The singular momentum in this case takes the value $k_s/(2\pi T) = - 3/40$, the modes were computed for momenta within the values $k\in (k_s+4\times10^{-4},-k_s)(2\pi T)$. The second quasinormal mode (right plot) does not show any peculiarity in its trajectory in the complex frequency plane. On the other hand, the purely imaginary pole (left plot), which in the hydrodynamic regime obeys the dispersion relation $\omega \approx -i \frac{\eta}{\epsilon+P} k^2$, deviates from the parabolic behavior when the momentum approaches $k_s$. In fact, the closer we go to the singular momentum, the fastest the imaginary part grows, in consistency with the presence of an instability at $k=k_s$. Therefore, this naive quasinormal mode analysis suggests a relation of the singular point with a possible instability of the RN black hole.
\para On the other hand, we would also like to emphasize that the presence of the characteristic surface is a consequence of being at finite density, i.e. the Schwartzshild black hole does not have a singular momentum (see appendix \ref{app:a}). That might be an indication that the RN black hole is not the real finite density groundstate of the system, and that the ``good'' background  would propagate fluctuations without any characteristic surface in the bulk.

\para Finally, let us make an analogy to the holographic dual of a two dimensional pure gravitational anomaly, which is described by the three dimensional Einstein gravity with a gravitational Chern-Simons term, i.e. the so called topologically massive gravity (TMG) \cite{Kraus:2005zm}. As it is well known anomaly-induced transport shows a lot of similarities between chiral four dimensional  quantum field theories and the two dimensional one \cite{Jensen:2012kj}. In particular, the mixed gauge-gravitational anomaly is the four dimensional cousin of the gravitational anomaly in two dimensions. The same is expected to be true for their holographic duals, i.e. the theory in consideration here (\ref{eq:action}) and TMG respectively. To make the analogies clearest let us enumerate them as follow
\begin{itemize}
	\item Both TMG and the gravitational theory (\ref{eq:action}) are higher derivatives in the metric sector and the equations of motion are third order. On the TMG side, a lot of research have been done with finite Chern-Simons coupling and it is believed that for some values of the coupling it is a consistent holographic theory, without other higher derivative terms or other matter fields coming from string theory. 
	
	In both cases  the Chern-Simons couplings have been studied non-perturbatively.   However, the consistency status is better understood in TMG,  in \cite{Li:2008dq} it was found that there always exist a bulk mode (either massless graviton or massive graviton) around AdS$_3$ with negative energy for generic Chern-Simons coupling except for the critical value, at which  appropriate boundary conditions were established and a holographic duality in terms of a logarithmic CFT in \cite{Grumiller:2008qz} was discussed. Subsequently it was conjectured that the stable backgrounds are given by the warped AdS$_3$ geometry \cite{Anninos:2008fx}. 
	
	\item Three and five dimensional Chern-Simons terms reproduce the right anomalous contribution in the Ward identities of a quantum field theory with either a pure gravitational anomaly in two dimensions or a mixed gauge-gravitational anomaly in four dimensions respectively.
	
	\item Due to the nature of the mixed gauge-gravitational Chern-Simons in 5-dimensions, either nontrivial gauge fields or finite temperature is needed in order to have a contribution into the equations of motion, in opposite with the TMG case, in which the Chern-Simons is purely gravitational. Therefore the analogous backgrounds to compare in the study should be AdS$_5$ black holes with AdS$_3$.
	
	The equations for the linearised graviton on AdS$_5$ are $\lambda$-independent, implying the stability and consistency of the linear problem as it happens with Einstein-Maxwell theory. At finite temperature but zero charge density, there is no signal of unstable modes in the five dimensional theory (see appendix \ref{app:a}), and no singular point emerges in the bulk.\footnote{However, an analysis of the absence (or existence) of negative energy states on this background is still missing.}

\end{itemize}

By the above analogy between TMG and the five dimensional gravitational theory (\ref{eq:action}) 
and the known studies on the consistency in TMG, 
one may expect that the five dimensional theory with finite Chern-Simons coupling, around Reissner-Nordstr\"om, is generically unstable but new stable groundstate could exist around which no puzzling singular momentum issue would emerge.

\section{Conclusion and discussion}
\label{sec5}
\para We have studied the perturbative instability for RN black hole in Einstein-Maxwell-Chern-Simons gravitational theory with a gravitational Chern-Simons term at both zero temperature and finite temperature. We have established that there exist two sources of instabilities. One is the BF bound violation on the AdS$_2$ near horizon geometry at zero temperature (similar to pure CS gauge theory case), and the resulting phase diagram at finite temperature is a bell curve which only  exist when $\alpha>{\alpha_c}_\text{min}$. Another instability is associated to $k_s$ and close to it there are always unstable modes. Moreover, this instability happens at arbitrary gravitational Chern-Simons coupling $\lambda$. At finite temperature we studied the origin of this singular momentum and found that it indicates whether a characteristic surface appears in the bulk or behind the horizon, in particular, when $k=k_s$ the singular point coincides with the horizon. The presence of such surface in the bulk turns unreliable the IR boundary value problem, maybe related to the presence of ghost-like modes and/or potentially causality issues.  
\para Thus our study suggests that the linearized problem of the theory with the gravitational Chern-Simons term around Reissner-Nordstr\"om is unreliable, for non-perturbative Chern-Simons coupling. However, considering the theory around AdS$_5$ is well defined, and the Schwartzshild black hole seems not to suffer of instabilities, at least at the quasinormal modes level.\footnote{Thus the calculations on the transport coefficients \cite{Landsteiner:2016stv,{Landsteiner:2011iq}} from this gravitational theory for zero density system are reliable. } It could be possible that for finite charge density a new well defined background may exist, as it is conjecture to happen in TMG with the case of AdS$_3$ and warp AdS$_3$. 

\para This project ended with several open questions that we hopefully will address in the future to properly understand the system. An incomplete list reads 
\begin{itemize}
\item Stability, causality and consistency analysis of the full theory, taking into account the possible presence of the Orstrogodski instability \cite{Woodard:2015zca},   positivity of time delay \cite{Camanho:2014apa} and the method of characteristics \cite{Reall:2014pwa}. 
\item Check the existence of a reliable new ground state at finite density for finite values of $\lambda$, and check its stability. 
\item Study the fully backreacted solution of the spatially modulated phase for the parameters inside the bell curve in Fig. \ref{fig:phase1}. 
\item Consider a consistent truncation of SUGRA including a gravitational Chern-Simons term \cite{Cremonini:2008tw, Myers:2009ij,Aharony:1999rz} and analyse the spatially modulated instability in this theory. 
\item A 4D gravitational analogue of the spatially modulated instability in 5D has been studied in the context of AdS/CFT in e.g. \cite{Donos:2011bh,{Bergman:2011rf}}. By including the effect of gravitational Chern-Simons term (e.g. \cite{Saremi:2011ab,Liu:2014gto}), one may attempt to expect that a similar instability will happen.  
\end{itemize}

\subsection*{Acknowledgments}
We thank R. G. Cai,  L. Y. Hung, K. Landsteiner, A. Marini, J. Ren, Y. W. Sun for helpful discussions. Y.L. has been supported by the Thousand Young Talents Program of China and grants ZG216S17A5 and KG12003301 from Beihang University. Y.L. thanks IFT-UAM/CSIC for the warm hospitality during the completion of the present work.  

\appendix
\section{QNMs for Schwartzshild black hole background}
\label{app:a}
\para
In this appendix, we shall show that the gravitational theory (\ref{eq:action}) with gravitational Chern-Simons term seems to be stable around the Schwartzschild black hole background, by the computation of the quasinormal modes spectra 
at zero density and finite or zero temperature.  
\para By studying the $\lambda$ effects on the spectrum of quasi-normal modes (QNMs) around the Schwartzschild black hole background, which encodes the locations of poles of the retarded two-point correlators of the energy-momentum and axial current operators, we will show the poles are in the lower half of the complex frequency plane. 
\para We consider the case of zero axial charge density. On the gravity side, we have the Schwartzschild AdS black hole 
\be
ds^2=-r^2 f(r) dt^2+\frac{dr^2}{r^2 f(r)}+r^2 (dx^2+dy^2+dz^2)\,,~~~ A=0\,,
\ee
with $f(r)=1-\frac{r_0^4}{r^4}$. 
We introduce the metric and gauge field fluctuations of form 
$\delta g_{\mu\nu}=h_{\mu\nu} e^{-i \omega t+ i k z},\delta A_\mu=a_{\mu} e^{-i \omega t+ i k z}.$ We work in the radial gauge $h_{r\mu}=a_r=0$ and classify the fluctuations according to their transformation properties in the $(x, y)-$plane as follows,  
\begin{itemize}
\item Helicity 2:  $h_{xy}, \frac{1}{2}(h_{xx}-h_{yy})$\,;
\item Helicity 1:  $h_{t\alpha}, h_{z\alpha}, a_\alpha$ with $\alpha\in (x,y)$\,;
\item Helicity 0:  $h_{tt}, h_{tz},h_{xx}+h_{yy}, h_{zz}, a_t, a_z$\,.
\end{itemize}
\para For both, helicity 2 and helicity 0 sectors, there is no effect from $\lambda$ on the linearised EOM and they satisfied the same equations as the Einstein-Maxwell gravity case. There is no instability from these sectors \cite{Kovtun:2005ev}. However, the gravitational anomaly term does have effects on the helicity 1 fluctuations. The independent equations are 
\bea
h''_{t\alpha}+\frac{1}{r}h'_{t\alpha}-\frac{k^2+4r^2 f}{r^4 f}h_{t\alpha}-\frac{k\omega}{r^4 f}h_{z\alpha}+12ik\lambda f'\epsilon^{\alpha\beta}\Big(a'_\beta-\frac{2}{r} a_\beta\Big)&=&0\,,\\
\frac{\omega}{f} h'_{t\alpha}-\frac{2\omega}{rf} h_{t\alpha}+k h'_{z\alpha}-\frac{2k}{r}h_{z\alpha}+\frac{12ik\lambda\omega f'}{f}\epsilon^{\alpha\beta}a_\beta&=&0\,,\\
a''_\beta+\Big(\frac{3}{r}+\frac{f'}{f}\Big)a'_\beta+\frac{1}{r^4 f}\Big(\frac{\omega^2}{f}-k^2\Big)a_\beta+12ik\lambda\frac{f'}{r^2f}\epsilon^{\beta\alpha}\Big(h'_{t\alpha}-\frac{2}{r}h_{t\alpha}\Big)&=&0\,.
\eea
where $\alpha\in (x, y)$ and $\epsilon^{xy}=-\epsilon^{yx}=1$.
\para  The gauge invariant quantities for the fluctuations are constructed as 
\be
Z_\alpha=k\frac{h_{t\alpha}}{r^2}+\omega \frac{h_{z\alpha}}{r^2}\,,~~~E_\alpha=\omega a_\alpha\,.
\ee
In terms of the gauge invariant fields the above fluctuation equations can be written as 
\bea
Z_x''+C_1 Z_x'+C_2 Z_x+C_3 E_y'+C_4 E_y&=&0\,,\nn\\
E_y''+D_1 E_y'+D_2 E_y+D_3 Z_x'&=&0\,,\nn\\
Z_y''+C_1 Z_y'+C_2 Z_y-C_3 E_x'-C_4 E_x&=&0\,,\nn\\
E_x''+D_1 E_x'+D_2 E_x-D_3 Z_y'&=&0\,,
\eea
where the coefficients read
\bea
C_1&=&\frac{5}{r}-\frac{\omega^2 }{k^2 f-\omega^2}\frac{f'}{f}\,,~~~~
C_2=-\frac{1}{r^4 f^2}\Big(k^2f-\omega^2\Big)\,,~~~~
C_3=12i\lambda\frac{k^2 f'}{\omega r^2}\,,\nn\\
C_4&=&-12i\lambda\frac{k^2f'}{\omega r^2 }\Big(\frac{2}{r}+\frac{\omega^2}{k^2 f-\omega^2}
\frac{f'}{f}\Big)\,,~~~~
D_1=\frac{3}{r}+\frac{f'}{f}\,,\nn\\
D_2&=&\frac{1}{r^4f^2}(\omega^2-k^2 f)+144\lambda^2\frac{k^2\omega^2 }{k^2 f-\omega^2}\frac{f'^2}{r^2f}\,,~~~~
D_3=-12i\lambda\frac{k^2\omega f'}{k^2 f-\omega^2}\,.\eea

\begin{figure}[h]
	\begin{center}
		\includegraphics[width=0.46\textwidth]{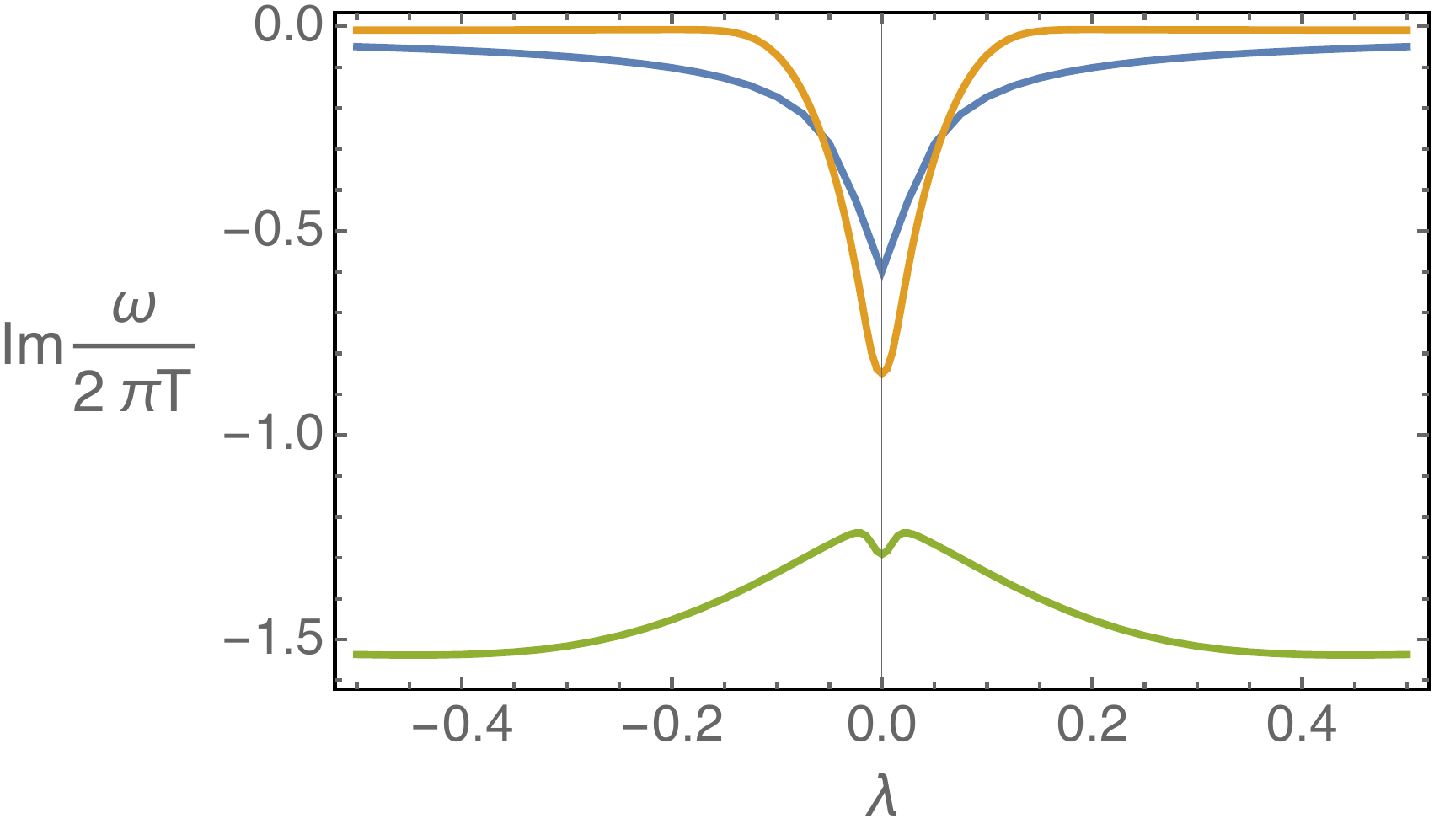}
		\includegraphics[width=0.46\textwidth]{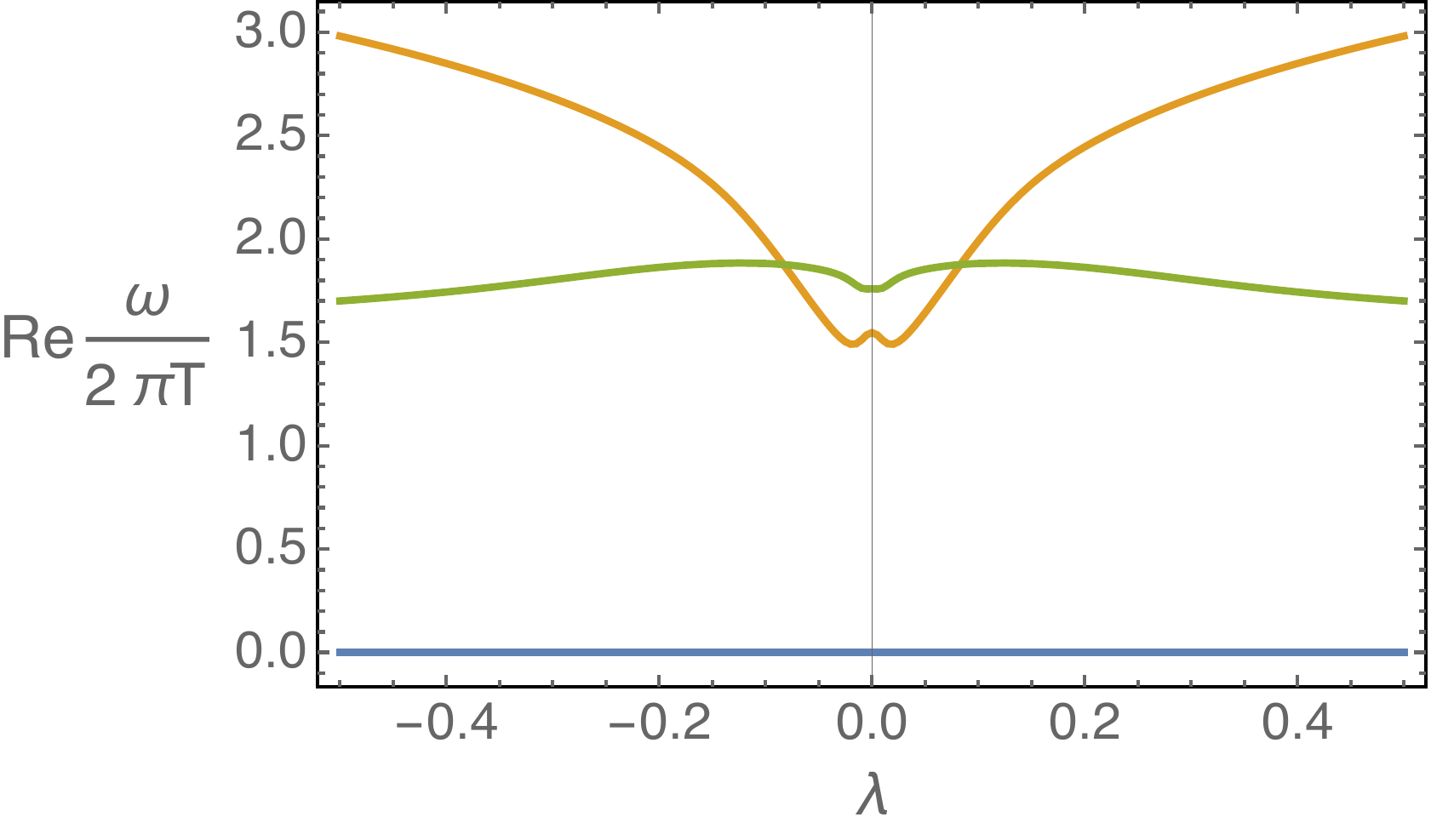}
		\caption{\small The imaginary and real parts of quasinormal frequency as a function of $\lambda$ at fixed $k=4\pi T$. The blue  line is for the Diffusion pole, the orange line is for the first QNMs and the green line is the second QNMs.}
		\label{fig:qnm}
	\end{center}
\end{figure}

\para Redefining the fields in term of 
$Z_\pm=Z_x\pm i Z_y, E_\pm=E_x\pm i E_y$, we have two set of decoupled equations given the helicity of the fields
\bea
\label{eq:zpm}
Z_\pm''+C_1 Z_\pm'+C_2 Z_\pm\pm (-i C_3) E_\pm'\pm (-iC_4) E_\pm&=&0\,,\\
\label{eq:epm}
E_\pm''+D_1 E_\pm'+D_2 E_\pm\pm (i D_3) Z_\pm'&=&0\,.
\eea
In order to solve for the QNM spectra for the above coupled differential equations, we expand the fields near the horizon
\be
Z_\pm=\Big(1-\frac{r_0^4}{r^4}\Big)^{-\frac{i\omega}{4r_0}}(z_{in}+...)\,,~~~~E_\pm=\Big(1-\frac{r_0^4}{r^4}\Big)^{-\frac{i\omega}{4r_0}}(e_{in}+...)\,,
\ee
and select the infalling modes. Near the AdS$_5$ boundary we have 
\be
Z_\pm=z_0(1+\dots)+\frac{z_1}{r^4}(1+\dots)\,,~~~~E_\pm=e_0(1+\dots)+\frac{e_1}{r^2}(1+\dots)\,.
\ee
\para The quasinormal frequencies are the values of complex $\omega$ for fixed real $k$, $z_0=e_0=0$. 
When $\omega/r_0, k/r_0\ll 1$, following \cite{Kovtun:2005ev} we can define $\zeta=\frac{\omega}{k}$, and solve the above equations peturbatively on $\zeta, k$. After doing so, we obtain the perturbative solutions in $\zeta, k$, 
which have a near boundary expansion
\bea
E_\pm&\simeq&\big(1-\frac{4ir_0\zeta}{k}\big)\beta+\mathcal{O}(k^2)+\mathcal{O}\big(\frac{1}{r^4}\big),\\
 Z_\pm &\simeq& \big(1-\frac{4ir_0\zeta}{k}\big)\alpha+\mathcal{O}(k^2)+\mathcal{O}\big(\frac{1}{r^2}\big).
 \eea
The normalizability condition then implies 
$\omega=-i \frac{k^2}{4r_0}$, which does not depend on $\lambda$. 
\para Next, we fixed $\frac{k}{4\pi T}=1$ and compute the three lowest quasinormal frequencies as a function of $\lambda$, we show them in figure \ref{fig:qnm}. The Re$\frac{\omega}{2r_0}$ is shown in the right plot and the Im$\frac{\omega}{2r_0}$ at the left. In the left plot we observe that the quasinormal modes are always located in the lower half plane. We also notice that quasinormal modes are symmetric under $\lambda\to -\lambda$. Thus we can claim that the linear problem around Schwartzschild is well defined, and no instabilities seem to appear at the quasinormal frequencies level.\footnote{One can do a similar analysis for QNM of the helical 1 sectors around Reinssner Nordstrom black hole.  Close to the singular momentum (\ref{eq:ksfinT}), the Diffusion pole (while not other QNM's) seems to cross the $\omega=0$ axis from a negative imaginary frequency to a positive imaginary frequency.}

\end{document}